\documentclass[a4paper,fleqn,amssymb,usenatbib,useAMS,usedcolumns]{mnras}
\usepackage{times} 
\usepackage{latexsym,amssymb,amsmath,amsfonts}
\usepackage{rotating} 
\usepackage{float}
\usepackage{graphicx}
\usepackage{natbib}
\usepackage{longtable}
\usepackage{url}
\usepackage{dcolumn}
\usepackage{textcomp}
\usepackage{multicol}
\usepackage{bm}
\usepackage{pdflscape}
\usepackage[T1]{fontenc}
\usepackage{ae,aecompl}
\usepackage[caption=false]{subfig}
\usepackage[dvipsnames]{xcolor}

\begin{document}

\newcommand{\cf}{{\textrm c.f.}}
\newcommand{\eg}{{\textrm e.g.}}
\newcommand{\ie}{{\textrm i.e.}}
\newcommand{\km}{\ensuremath{\mbox{~km}}}
\newcommand{\pc}{\ensuremath{\mbox{~pc}}}
\newcommand{\kpc}{\ensuremath{\mbox{~kpc}}}
\newcommand{\ebv}{\mbox{$E(B-V)$}}
\newcommand{\bvri}{\mbox{$BVRI$}}
\newcommand{\degree}{\mbox{$^\circ$}}
\newcommand{\maghundred}{\mbox{mag (100 d)$ ^{-1} $}}
\newcommand{\msun}{\mbox{M$_{\odot}$}}
\newcommand{\msol}{\mbox{M$_{\odot}$}}
\newcommand{\zsol}{\mbox{Z$_{\odot}$}}
\newcommand{\rsun}{\mbox{R$_{\odot}$}}
\newcommand{\kms}{\mbox{$\rm{\,km\,s^{-1}}$}}
\newcommand{\kkms}{\mbox{$\times10^3\rm{\,km\,s^{-1}}$}}
\newcommand{\ergs}{\mbox{$\rm{\,erg\,s^{-1}}$}}
\newcommand{\logl}{\mbox{$\log L/{\rm L_{\odot}}$}}
\newcommand{\nickel}{\mbox{$^{56}$Ni}}
\newcommand{\cobalt}{\mbox{$^{56}$Co}}
\newcommand{\iron}{\mbox{$^{56}$Fe}}
\newcommand{\mum}{\mbox{$\mu{\rm m}$}}
\newcommand{\el}{\mbox{${e}^{-}$}}
\newcommand{\ld}{\mbox{$\lambda$}}
\newcommand{\ldld}{\mbox{$\lambda\lambda$}}
\newcommand{\ergscm}{{$ \mathrm{erg\ s^{-1} cm^{-2}}$}}
\newcommand{\ergscma}{{$ \mathrm{erg\ s^{-1} cm^{-2} arcsec^{-2}}$}}
\newcommand{\fcgs}{{$ \mathrm{erg\ s^{-1} cm^{-2}}$\AA$^{-1}$}}
\newcommand{\arc}{$\mathrm{^{\prime\prime}}$}
\newcommand{\cross}{$\mathrm{\times}$}
\newcommand{\apr}{$\mathrm{\approx}$}
\newcommand{\til}{$\mathrm{\sim}$}
\newcommand{\zem}{$\mathrm{$z_{em}$}$}
\newcommand{\plm}{$\mathrm{\pm}$}
\newcommand{\sblya}{$\mathrm{SB_{Ly\alpha}}$}
\newcommand{\hi}{\mbox{H\,{\sc i}}}
\newcommand{\mgii}{\mbox{Mg\,{\sc ii}}}
\newcommand{\mgi}{\mbox{Mg\,{\sc i}}}
\newcommand{\feii}{\mbox{Fe\,{\sc ii}}}
\newcommand{\oi}{\mbox{O\,{\sc i}}}
\newcommand{\cii}{\mbox{C\,{\sc ii}}}
\newcommand{\ci}{\mbox{C\,{\sc i}}}
\newcommand{\sii}{\mbox{Si\,{\sc ii}}}
\newcommand{\znii}{\mbox{Zn~{\sc ii}}}
\newcommand{\caii}{\mbox{Ca\,{\sc ii}}}
\newcommand{\nai}{\mbox{Na\,{\sc i}}}
\newcommand{\civ}{\mbox{C\,{\sc iv}}}
\newcommand{\heii}{\mbox{He\,{\sc ii}}}
\newcommand{\nv}{\mbox{N\,{\sc v}}}
\newcommand{\siv}{\mbox{Si\,{\sc iv}}}
\newcommand{\flya}{$f\mathrm{_{Ly\alpha}}$}
\newcommand{\llya}{$L\mathrm{_{Ly\alpha}}$}
\newcommand{\fciv}{$f\mathrm{_{CIV}}$}
\newcommand{\fheii}{$f\mathrm{_{HeII}}$}
\newcommand{\fnv}{$f\mathrm{_{NV}}$}
\def\h2{$\rm H_2$}
\def\Nh2{$N$(H${_2}$)}
\def\chin{$\chi^2_{\nu}$}
\def\chiu{$\chi_{\rm UV}$}
\def\sys{J0441$-$4313~}
\def\lya{\ensuremath{{\rm Ly}\alpha}}
\def\lymana{\ensuremath{{\rm Lyman}-\alpha}}
\def\kms{km\,s$^{-1}$}
\def\cms{cm$^{-2}$}
\def\cc{cm$^{-3}$}
\def\zabs{$z_{\rm abs}$}
\def\zem{$z_{\rm em}$}
\def\nhi{$N$($\hi$)}
\def\ln{log~$N$}
\def\nh{$n_{\rm H}$}
\def\ne{$n_{e}$}
\def\21{21-cm}
\def\ts{T$_{s}$}
\def\th{T$_{01}$}
\def\t0{T$_{0}$}
\def\ll{$\lambda\lambda$}
\def\l{$\lambda$}
\def\fc{$C_{f}$}
\def\c21{$C_{21}$}
\def\mjb{mJy~beam$^{-1}$}
\def\taudv{$\int\tau dv$}
\def\taup{$\tau_{\rm p}$}
\def\ha{H\,$\alpha$}
\def\hb{H\,$\beta$}
\def\oi{[O\,{\sc i}]}
\def\oii{[O\,{\sc ii}]}
\def\oiii{[O\,{\sc iii}]}
\def\nii{[N\,{\sc ii}]}
\def\sii{[S\,{\sc ii}]}
\def\taudvl{$\int\tau dv^{3\sigma}_{10}$}
\def\taudv{$\int\tau dv$}
\def\vshift{$v_{\rm shift}$}
\def\wmg{$W_{\mgii}$}
\def\wfe{$W_{\feii}$}
\def\dgi{$\Delta (g-i)$}
\def\ebv{$E(B-V)$}

\def\sig{$\sigma$}
\def\pathfig{images/}

%

\title[Ly$\alpha$ emission from M151304.72-252439.70]{
{Lyman-$\alpha$ emission from a WISE-selected optically faint powerful radio galaxy M151304.72-252439.7 at $z$ = 3.132 
}\thanks{Based on observations made with the Southern African Large Telescope (SALT).}}

\author[Shukla et.al]{Gitika Shukla$^{1}$\thanks{E-mail: gitika@iucaa.in}, Raghunathan Srianand$^{1}$, Neeraj Gupta$^{1}$, Patrick Petitjean$^2$,
\newauthor{Andrew J. Baker$^3$, Jens-Kristian Krogager$^2$, Pasquier Noterdaeme$^2$}
{} \\
\\
 $^{1}$Inter-University Centre for Astronomy and Astrophysics (IUCAA), Post Bag 4, Pune 411007, India \\
 $^{2}$Institut dAstrophysique de Paris, UMR 7095, CNRS-SU, 98bis boulevard Arago, 75014 Paris, France\\
 $^{3}$Department of Physics and Astronomy, Rutgers, the State University of New Jersey, 136 Frelinghuysen Road, Piscataway,\\ NJ 08854-8019, USA\\
}

\date{ }
\pubyear{}
\maketitle
\label{firstpage}
\pagerange{\pageref{firstpage}--\pageref{lastpage}}
%
%
\begin {abstract}  
\par\noindent
We report the detection of a large ($\sim90$ kpc) and luminous \lya\ nebula [\llya\ = (6.80\plm0.08) $\times 10^{44}$ \ergs] around an optically faint (r$>23$ mag) radio galaxy M1513-2524  at \zem=3.132. The double-lobed radio emission has an extent of 184 kpc, but the radio core, i.e., emission associated with the active galactic nucleus (AGN) itself, is barely detected. This object was found as part of our survey to identify high-$z$ quasars based on Wide-field Infrared Survey Explorer (WISE) colors. The optical spectrum has revealed \lya, \nv, \civ\ and \heii\ emission lines with a very weak continuum. Based on long-slit spectroscopy and narrow band imaging centered on the \lya\ emission, we identify two spatial components: a ``compact component" with high velocity dispersion ($\sim 1500$ \kms) seen in all three lines, and an ``extended component", having low velocity dispersion (i.e., 700-1000 \kms). The emission line ratios are consistent with the compact component being in photoionization equilibrium with an AGN. We also detect spatially extended associated \lya\ absorption, which is blue-shifted within 250-400 \kms\ of the \lya\ peak. The probability of \lya\ absorption detection in such large radio sources is found to be low ($\sim$10\%) in the literature. M1513-2524 belongs to the top few percent of the population in terms of \lya\ and radio luminosities. Deep integral field spectroscopy is essential for probing this interesting source and its surroundings in more detail. 
\end{abstract}

%
%
%
\section{Introduction} 
\label{sec_introduction}
The ubiquitous  presence  of extended \lya\ emission around diverse populations of galaxies, ranging from quasars \citep{heckman1991a,heckman1991b,Borisova2016,Arrigoni2019} to powerful high-\emph{z} radio galaxies (HzRGs) \citep[see][]{chambers1990,villar2003,villar2007} and several other populations such as sub-millimetre and Lyman-break galaxies \citep{matsuda2004,chapman2001,geach2014}, is now well established. The detection rate of such diffuse \lya\ emission from high-$z$ galaxies has remarkably gone up to 100$\%$, in some recent studies \citep{Borisova2016,Arrigoni2019}, using integral field spectrographs like the Multi-Unit Spectroscopic Explorer \citep[MUSE;][]{bacon2010} and the Keck Cosmic Web Imager \citep[KCWI;][]{morrissey2012} on 8-10 metre class telescopes, operating at excellent astronomical sites. In particular, these instruments have increased the detection of faint and small scale \lya\ nebulae \citep{Hu1987,farina2017,wisotzki2018}. In principle, these extended \lya\ emitting regions provide important means to study the properties of the circumgalactic medium (CGM) and the interface between CGM and the intergalactic medium (IGM) around high-$z$ galaxies. Detailed studies of the spatial distribution, kinematics and excitation of the gas traced by the extended \lya\ emission can provide vital clues on various feedback processes that drive star formation in high-$z$ galaxies.
\vskip 0.1cm

Traditionally, studies of the CGM and IGM have been carried out using absorption lines detected against bright background sources, typically bright quasars. The one-dimensional nature of absorption line studies, however, is not ideal for probing the spatial distribution of the gas surrounding individual galaxies. Despite several studies dedicated to understand the complex interactions between gas surrounding the host galaxies and the galaxies themselves, very little progress has been made to date. In order to fully characterise the physical and kinematic properties of this gas, it is important to combine absorption and emission studies.

The recent detections of enormous \lya\ nebulae \citep[ELANs;][]{matsuda2004,Cantalupo2014,cai2017,cai2018,Arrigoni2018,Arrigoni2019}, which are characterized by very extended \lya\ halos (\lya\ extent $>200$ kpc) with luminosity \llya\ $>10^{44}$\ergs\ and surface brightness \sblya\ $>10^{-17}$\ergscma, have drawn a lots of attention. As the diffuse emission in these cases extend well beyond the virial radii of the galaxies, ELANs can be a very good probe of the dark matter potential of the halos hosting the galaxies and the interface regions between CGM and IGM.  However, ELANs are rare, and only 1\% of quasars seem to posses them. Based on their radial surface brightness profiles, it has been proposed that ELANs trace regions of higher volume density and/or have additional sources of ionizing radiation embedded in their halos \citep{Cantalupo2014,Arrigoni2018}. However, a better understanding of these scenarios requires further investigation.

The extended \lya\ emission studies centred on HzRGs and radio-loud quasars have shown a clear correlation between powerful radio jets and diffuse gas \citep{vanojik1997,heckman1991a,heckman1991b,villar2007}, where the jet clearly seems to alter the kinematics and morphology of the \lya\ emitting gas. In a sample of 18 $z>2$ radio galaxies, \citet{vanojik1997} have found a clear correlation between radio source size and the size of the diffuse \lya\ emission, along with an anti-correlation between \lya\ velocity width and radio size. In more than 60\% of these cases, strong associated \hi\ absorption ($N_\mathrm{{HI}}\geq 10^{18} \mathrm{cm^{-2}}$) was detected. The detection rate of \hi\ absorption was found to be higher in smaller radio sources (i.e.,  $\sim 90\%$ when the source size is $<50$ kpc and 25\% when $>50$ kpc). In 61$\%$ of the cases, \lya\ was more extended than the radio source itself. The inner parts of the \lya\ halo within the extent of the radio emission showed perturbed kinematics (FWHM $>1000$ \kms ) due to jet-gas interaction, whereas the more extended (\til $100$ kpc) regions were dominated by quiescent kinematics (FWHM $<700$ \kms ). These results demonstrate that HzRGs reside in gas rich environments. Deep IR imaging studies have also revealed the presence of galaxy proto-clusters around high-$z$ radio galaxies \citep[][]{Mayo2012,Galametz2012,Wylezalek2013,Dannerbauer2014}. These proto-clusters show presence of high density environments around high-$z$ radio galaxies.

Several possible physical mechanisms have been proposed to power the extended \lya\ emission. In many cases the \lya\ nebulae are found to be associated with highly obscured type-II AGNs \cite[see, e.g.,][]{dey2005,bridge2013,overzier2013,hennawi2015,ao2017}. The most commonly discussed origins of \lya\ emission are:  (i) shock induced radiation, powered by radio jets or outflows \citep{mori2004,allen2008}; (ii) gravitational cooling radiation/ \lya\ collisional excitation \citep{haiman2000,dijkstra2006,Rosdahl2012}; (iii) fluorescent \lya\ emission due to photoionization by UV luminous sources like AGNs or star forming galaxies \citep{mccarthy1993,cantalupo2005,geach2009,overzier2013} and (iv) resonant scattering of \lya\ photons from embedded sources \citep{villar1996,dijkstra2008}. The presence of high-ionization lines like \civ\ and \heii\ can provide additional information on the kinematics of the gas and help disentangle the various physical processes powering the \lya\ emission. Extended emission in these high ionization lines is observed for a few HzRGs over tens of kpc \citep{villar2003,Morais2017}. There are a few detections in quasars as well, which have shown extended emission in these lines out to hundreds of kpc \citep[][]{marques2019}. \\
For the recently detected ELANs, photoionization by associated bright quasars has been suggested as the dominant powering source for the \lya\ emission \citep{Borisova2016}. However, contributions from other mechanisms like resonant scattering cannot be entirely ruled out \citep{Arrigoni2019}. In the case of HzRGs showing \lya\ extending beyond the associated radio emission, jet-gas interaction is seen to play a vital role in determining the properties of the \lya\ nebulae; however, to produce \lya\ on spatial scales larger than that of the radio emission, photoionization by a continuum source is still required \citep{vanojik1997}. Stellar photoionization has also been shown to increase the ratios \flya/\fciv\ and \flya/\fheii, especially for ``\lya-excess'' objects. There are cases where \lya\ nebulae are detected around radio-quiet type-II AGNs \citep{Prescott2015,marques2019}. \cite{Nelsson2006} had reported the presence of a 60 kpc large \lya\ nebula not associated with any optical source, which was argued to be powered by gravitational cooling radiation. However, later \cite{Prescott2015} argued that the absence of continuum at optical and longer wavelengths does not necessarily indicate gravitational origin for \lya\ emission. If anything, gravitationally powered \lya\ emission should instead be accompanied by a galaxy forming at the center of the \lya\ nebula, whose stellar continuum one should be able to detect.

\begin{figure}
            
           \includegraphics[viewport=60 20 800 520,width=10cm,height=7cm clip=true]{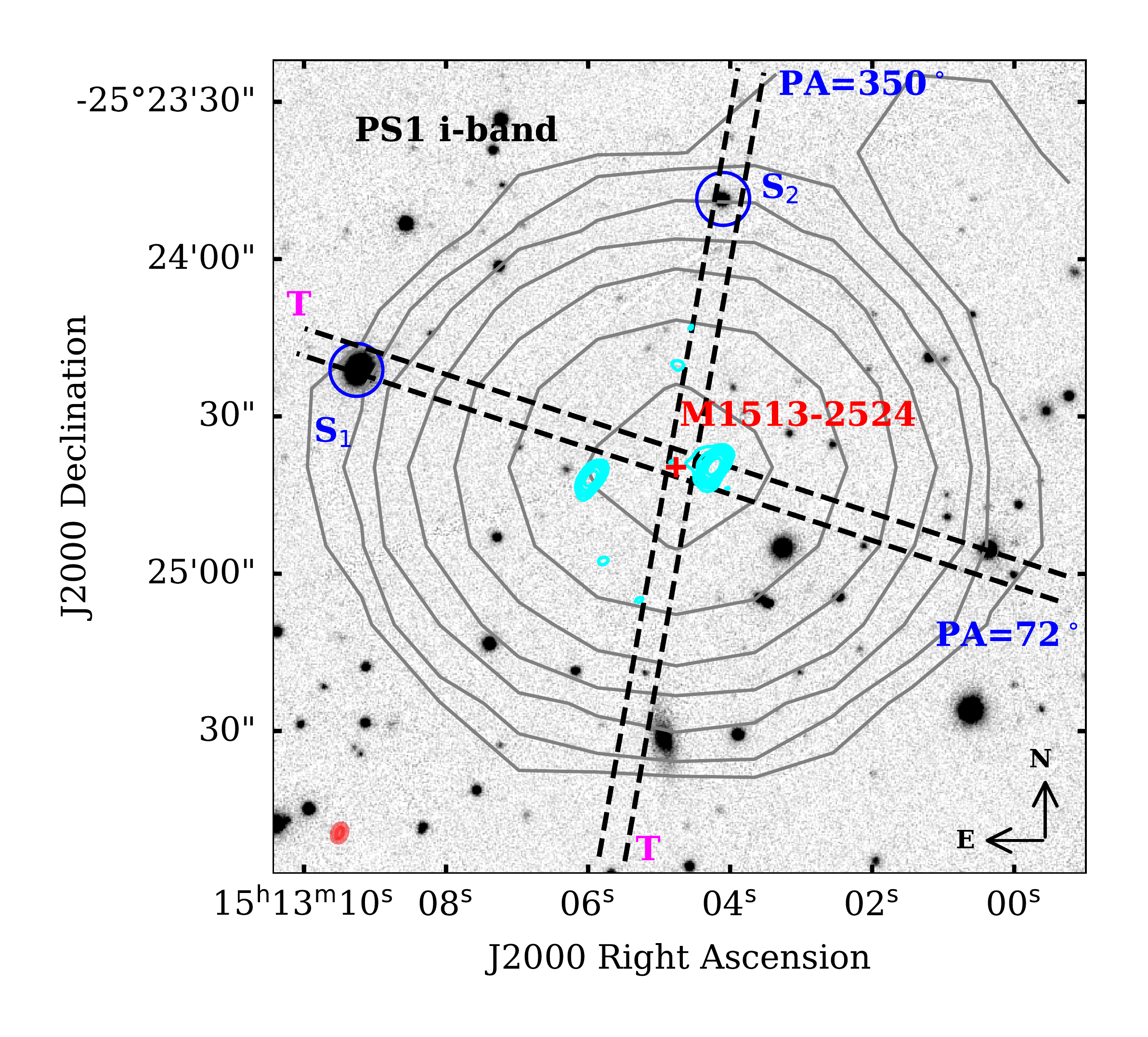}
	\caption{The i-band image of the field of M1513-2524 obtained in Pan-STARRS1. The two slit positions (PA=72\degree\ and PA=350\degree) along which the long-slit observations were carried out with respect to reference stars (identified as S$_1$ and S$_2$) falling within the slit are shown using dashed lines. The red plus marks the WISE position (RA=$\mathrm{15^h 13^m 04.72^s}$, Dec= $\mathrm{-25^d 24^m 39.70^s}$) where no optical source is detected. The cyan contours correspond to Band-5 (1.4 GHz) radio emission from our uGMRT observations, and the corresponding synthesized beam is shown in red in the lower left corner. The contour levels are plotted at 1.6 \cross\ (-1, 1, 2, 4, 8, 16, 32, 64, ..) $\mathrm{mJy\ beam^{-1}}$. Radio contours from NVSS are also over plotted (gray color). These contour levels are at 2.1 \cross\ (-1, 1, 2, 4, 8, 16, 32, 64, ..) $\mathrm{mJy\ beam^{-1}}$. The symbol ``T" corresponds to top of the 2D long-slit spectra shown in Figs. \ref{fig_2dspectrum} and \ref{fig_2dspectrum_sub}.}
         \label{fig_ps1}   
        \end{figure}

We have recently completed deep long-slit spectroscopic observations of extended \lya\ emission in 25 newly discovered radio bright (1.4 GHz flux density in excess of 200 mJy) quasars and radio galaxies at $z\ge2.7$ using the Southern African Large Telescope (SALT). These objects were discovered by us, as part of target selection for the MeerKAT Absorption Line Survey \citep[MALS;][]{Mals2,Mals3}. Here, we present a detailed analysis of the extended \lya\ emission around a special radio source, M151304.72-252439.70 (where M stands for MALS), in our sample. In the literature, this radio source is also known as MRC1510-252 and TXS1510-252. The source not only has higher values of radio power, angular size and \lya\ luminosity than other objects in our sample but also than other known radio sources at these redshifts. Interestingly, an optical counterpart is not detected in any of the optical images from PanSTARRS-1 (PS1) \citep{ps12018}, and only a very faint source is detected in infrared images from the Wide-field Infrared Survey Explorer (WISE) (see Fig.~\ref{fig_ps1}). Our survey spectrum clearly revealed the strong \lya, N~{\sc v}, C~{\sc iv} and He~{\sc ii} emission lines but with a very faint continuum emission.  The \lya\ emission is found to be extended with a clear signature of associated absorption. All these attributes make M1513-2524 an interesting object for detailed investigation.

We have organized this paper as follows. In Section \ref{sec_discover_follow_up} we have provided the details of our long-slit spectroscopic and narrow-band imaging observations using SALT, along with upgraded Giant Meterwave Radio Telescope (uGMRT) observations and data reduction. In Section \ref{sec_results}, we present the main results of our study; in particular, we explore the  connection between the radio jets/lobes and \lya\ emitting gas. We compare the \lya\ flux and radio size of M1513-2524 with those of radio sources at similar redshifts in the literature.
We discuss the properties of the associated \lya\ absorption and the detection of extended \civ\ and \heii\ emission lines in this section. We also provide a list of candidate \lya\ emitters found in our narrow band observations. In Section \ref{sec_summary}, we summarize and discuss our results. Throughout this paper, we have adopted a cosmology with H$_0$ = 67.4 \kms Mpc$^{-1}$, $\mathrm{\Omega_m}$ = 0.315 and  $\mathrm{\Omega_\Lambda}$ = 0.685 \citep[see][]{planck2018}. At the emission redshift (\emph{z} \til\ 3.13) for this cosmology, 1\arc\ corresponds to 7.8 kpc.
%
\section{observations and data reduction}
\label{sec_discover_follow_up}
In this section, we present details of the acquisition and analysis of various data used in this work.

\begin{table*}
\caption{Log of RSS/SALT long-slit observations for M1513-2524$^+$}
\begin{tabular}{cccccccc}
\hline
\hline
PA    & Observing date  & Wavelength coverage& FWHM of SPSF & Air mass & \multicolumn{1}{c}{Exposure time(s)} & \\
(deg) &     year/mm/dd  &(\AA)     &  (arcsec)  &          &        & \\  
\hline
 $72$ &2016-02-17 &4486-7533 &1.74 &1.19 & 2$\times$570 &  \\
 " &2017-05-23 &" &1.77 &1.38 & 2$\times$1200 &  \\
 $350$&2017-05-22 &" &1.38 &1.36 & 2$\times$1200 &  \\
\hline

\end{tabular}\\
\begin{flushleft}
$^+$ Slit width of 1.5 arcsec was used.\\
\end{flushleft}
\label{tab_observations}
\end{table*}

\subsection{Long-slit observations with RSS/SALT}
\label{sub_longslit}

To obtain the optical spectrum, we used the Robert Stobie Spectrograph (RSS) \citep{burgh2003,kobul2003} on the Southern African Large Telescope (SALT) \citep{buckley2006} in long-slit mode (Program IDs: 2016-2-SCI-017, 2017-1-SCI-016). The primary mirror of SALT is 11m across, consisting of 91 1m individual hexagonal mirrors. The RSS consists of an array of three CCD detectors with 3172 \cross\ 2052 pixels. We have used 2 \cross\ 2 pixel binning to improve the signal-to-noise ratio (SNR). These observations were carried out in service mode using a long-slit of width 1.5\arc, grating PG0900, GR-ANGLE=15.875$^o$ and CAMANG=31.75$^o$. This combination provides a wavelength coverage of 4486-7533 \AA, excluding the wavelength ranges 5497-5551 \AA\ and 6542-6589 \AA\ that correspond to gaps between CCDs. These settings were chosen such that \lya, \civ\ and \heii\ emission from the radio source are simultaneously covered and  Ly$\alpha$ emission falls in the most sensitive part of the spectrograph. The spectral resolution achieved is in the range of 200-300 $\mathrm{km\ s^{-1}}$. \\

\begin{figure*}
 \centering
    \includegraphics[viewport=20 35 2000 680,width=20cm, clip=true]{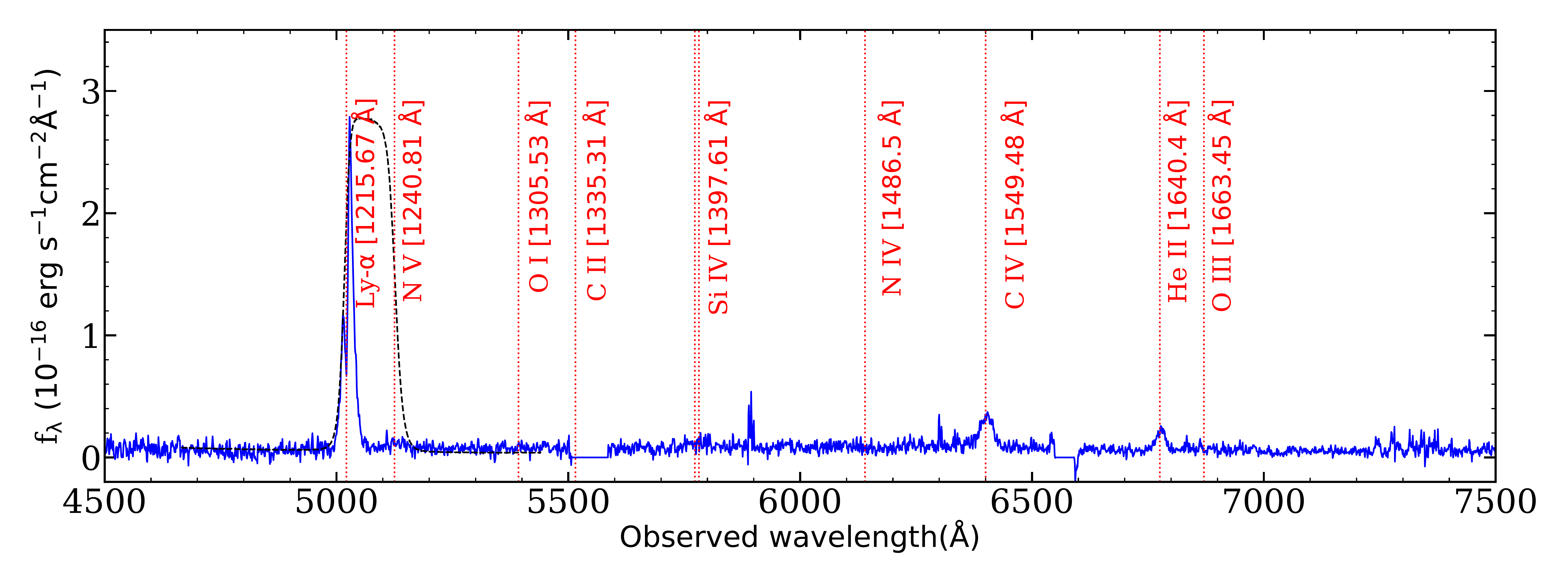}
   \caption{1D spectrum of M1513-2524 obtained using SALT. The \lya, \civ, \heii\ and weak \nv\ emission lines are detected, while other emission lines like \oi\  and \siv\  were not detected. The response curve of the filter ``PI05060'' used for the narrow band imaging covering \lya\ emission is shown as a black dashed curve. Vertical dotted lines mark the expected locations of commonly seen emission lines in AGNs \citep[see][]{vernet2001}.
   }
         \label{fig_1dspectrum}   
\end{figure*}

During the survey, the source was observed using a long-slit oriented at a position angle (PA) of 72\degree. To better understand the spatial distribution of gas, we followed up the target with deeper long-slit observations, keeping the slit oriented along PAs of 72\degree\ and 350\degree. These PAs were chosen so that we could have a blind off-set pointing at the WISE location of the source using a bright star's location (see Fig. \ref{fig_ps1}). The total on-source exposure times were 2400 s for each PA, split into 2 exposures of 1200 s each, with a dither of 2\arc\ along the slit. Seeing measured from the profiles of stars in the acquisition image are typically in the range 1.6\arc\ to 2.0\arc\ during our observations. The spectral point spread function (SPSF) for each PA was constructed from spatial profiles extracted after collapsing a region between 5000-5020 \AA\ for the reference stars. We measured SPSFs of 1.7\arc\ and 1.4\arc\ for spectra obtained along PA = 72\degree\ and 350\degree, respectively (see Table \ref{tab_observations}). The data were reduced using the SALT science pipeline \citep{crawford2010} and standard {\tt IRAF} tasks\footnote{IRAF is distributed by the National Optical Astronomy Observatories, which are operated by the Association of Universities for Research in Astronomy, Inc., under cooperative agreement with the National Science Foundation.}. For cosmic ray removal, we used the {\tt IRAF} based algorithm proposed by \cite{vandokkum2001}. The wavelength calibration was performed using a Xenon arc lamp. The spectrophotometric standard star EG21 (RA=$\mathrm{03^h 10^m 30.98^s}$, Dec= $\mathrm{-68^d 36^m 02.20^s}$) was used for flux calibration, where corrections for atmospheric extinction and air mass have been taken into account. The wavelengths were then shifted to  vacuum wavelengths. The final combined 1D spectrum of M1513-2524 is shown in Fig.~\ref{fig_1dspectrum}.

\begin{figure} 
           
             \includegraphics[viewport=45 370 1000 800,width=18cm,height=8cm, clip=true]{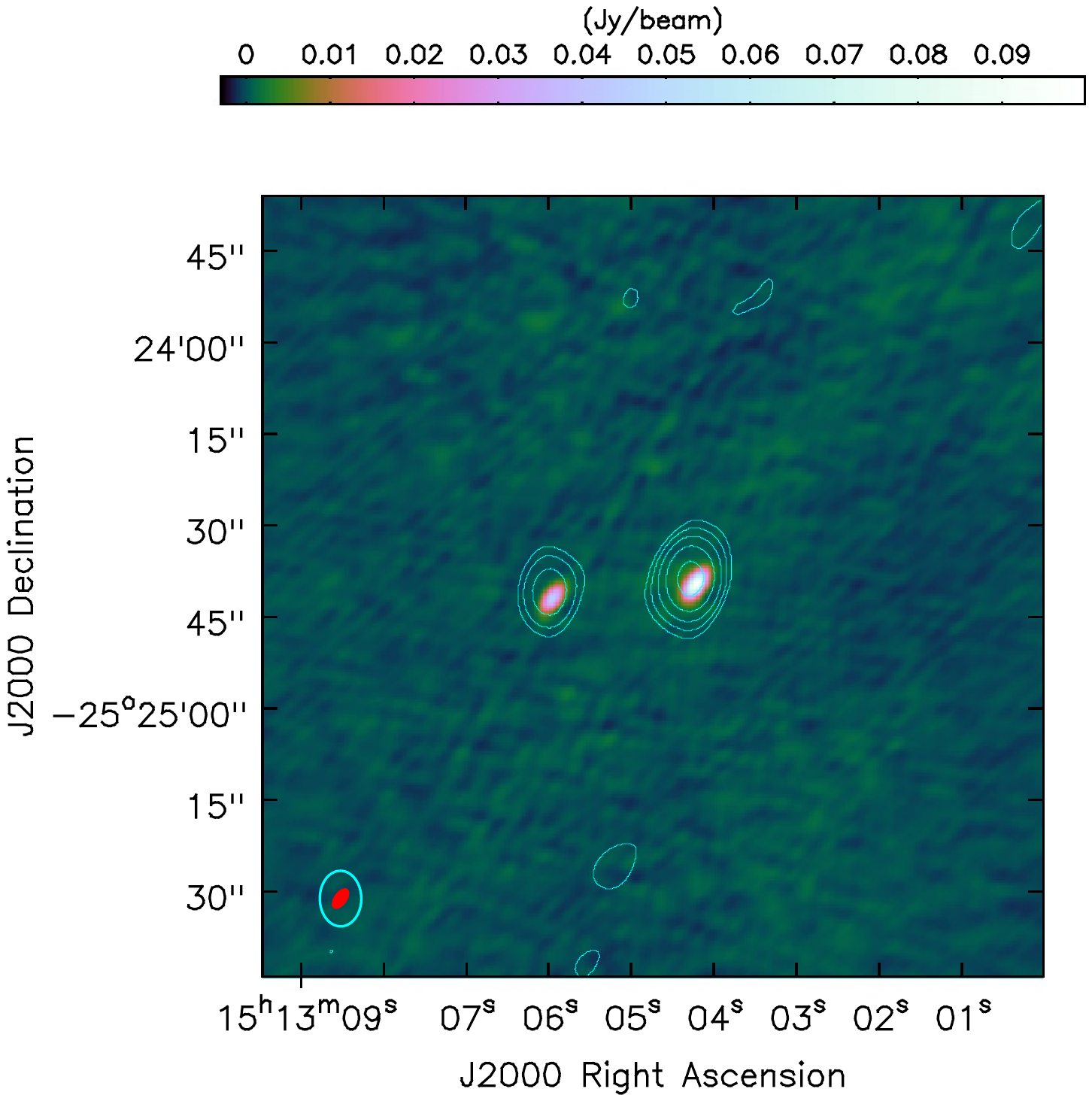}
            \caption{  
            uGMRT Band-3 (420 MHz; rms = 10 $\mathrm{mJy\ beam^{-1}}$) contours overlaid on Band-5 (1360 MHz; rms=0.4 $\mathrm{mJy\ beam^{-1}}$) image of M1513-2524. The synthesized beams provided in the text are plotted in the lower left corner. The color wedge at the top corresponds to the Band-5 image. The contour levels are:  40 $\times$ (-1, 1, 2, 4, 16, 32, 64, ...) $\mathrm{mJy\ beam^{-1}}$.}
            
         \label{fig_gmrt}   
        \end{figure}
\subsection{uGMRT observations}
\label{sub_gmrt}
The uGMRT Band-5 (1050-1450 MHz) and Band-3 (250 - 500 MHz) observations of M1513-2524 were carried out as part of our larger surveys to map the radio continuum and search for \hi\ 21-cm absorption associated with WISE selected quasars for MALS. For the Band-5 observations, the uGMRT wideband correlator was configured to cover 1260-1460 MHz. For Band-3, a frequency range of 300-500 MHz was covered. The on-source observing times for M1513-2524 were 10 mins for Band-5 and  30 mins for Band-3 observations. In both cases, only parallel hand correlations were recorded and standard calibrators for flux density, delay and bandpass calibrations were observed. These data were reduced using the Automated Radio Telescope Imaging Pipeline (ARTIP) that has been developed to perform the end-to-end processing (i.e., from the ingestion of the raw visibility data to the spectral line imaging) of data from the uGMRT and MeerKAT absorption line surveys. The pipeline is written using standard Python libraries and the Common Astronomy Software Applications (CASA) package; details are provided in \cite{Neeraj2020}. In short, following data ingestion, the pipeline automatically identifies bad antennas, baselines, time ranges and radio frequency interference  (RFI), using directional and median absolute deviation (MAD) statistics. After excluding these bad data, the complex antenna gains as a function of time and frequency are determined using the standard flux/bandpass and phase calibrators. Applying these gains, a continuum map is obtained, which is then self-calibrated. The uGMRT continuum maps obtained in Band-3 and Band-5 are shown in Fig. \ref{fig_gmrt}. The synthesized beams of the final images are 9.1\arc $\times$6.8\arc\ with PA=1.0\degree\ and 3.7\arc $\times$1.9\arc\ with PA=-35.5\degree, respectively. The flux density measurements from our uGMRT observations towards different lobes are summarized in Table~\ref{tab_radio}. Total flux density measurements from the literature are summarized in Table~\ref{tab_photometry}.

\subsection{Narrow-band imaging observations with RSS/SALT}
\label{sub_NB}
In order to accurately determine the location of the optical source and quantify any influence of the radio source on the overall distribution of \lya\ emitting gas, we obtained deep narrow band images of M1513-2524 using RSS filter PI05060, centered at  $\lambda_{cen} \approx 5071.5$ \AA\ with an FWHM of 110.5 \AA\ (see Fig. \ref{fig_1dspectrum} for the response curve of the  filter). The data were registered using 1\cross1 binning with a pixel scale of 0.1267\arc\ per pixel. Each individual frame is a combination of 3 CCD chips and 2 CCD gaps with a circular field of view of $(7.32^\prime)^2$. The observations were carried out such that the WISE source lay on the central CCD chip, and individual frames were dithered to avoid CCD artefacts. The observations were performed in service mode on March 2, 2019 with total exposure time of 2400 s, split into 4 exposures of 600 s each. During the course of the observations, 25\arc\ of dither was adopted between exposures. The data were first corrected for bias, cross talk and gain. We further used standard {\tt IRAF} tasks to remove cosmic rays and flat fielded the individual frames using dome flats. 

To perform the astrometric calibration of the narrow band images, we used 22 stars distributed across the central CCD chip,  for which accurate coordinates can be obtained from the PS1 survey. After astrometric calibration, objects in our narrow band image have a positional accuracy of 0.14\arc. As a next step, we subtracted the background from each individual exposure. To do that, we have utilised our script written in \textsc{Python}, employing the background subtraction technique provided by \textsc{Astropy}, a commumity developed \textsc{Python} package for astronomy \citep{astropy1,astropy2}. In this code, we essentially mask all $>$ 2\sig\ sources and the CCD gap regions and use the remaining pixels to estimate a 2D background model. We used \textsc{SExtractorBackground} as a background estimator (see the \textsc{Background2D} class of \textsc{Photutils} for details) in this process. We also performed the subtraction using the \textsc{Medianbackground} estimator, and the results from the two methods were consistent within 5$\%$. After subtracting the 2D background model, we subtracted further a constant mean background estimated after masking all the sources above 2\sig. Finally, we combined the narrow band images using the {\tt IRAF} task ``{\tt imcombine}", with all the images scaled and weighted by the ``median" counts measured in each frame.

\begin{figure*}

            \includegraphics[viewport=30 10 720 800,width=8.5cm,height=10cm, clip=true]{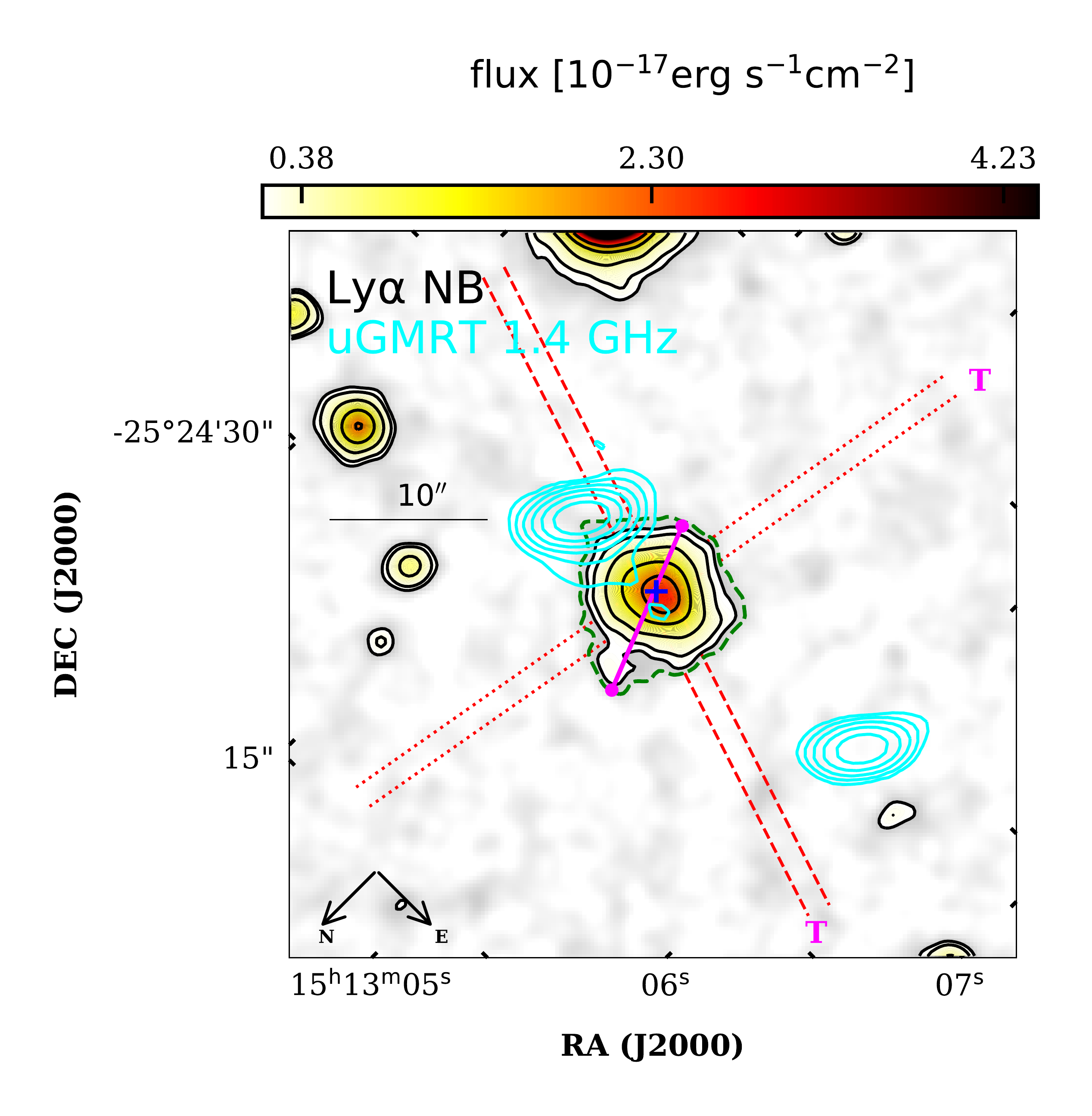}
            \includegraphics[viewport=30 10 720 800,width=8.5cm,height=10cm, clip=true]{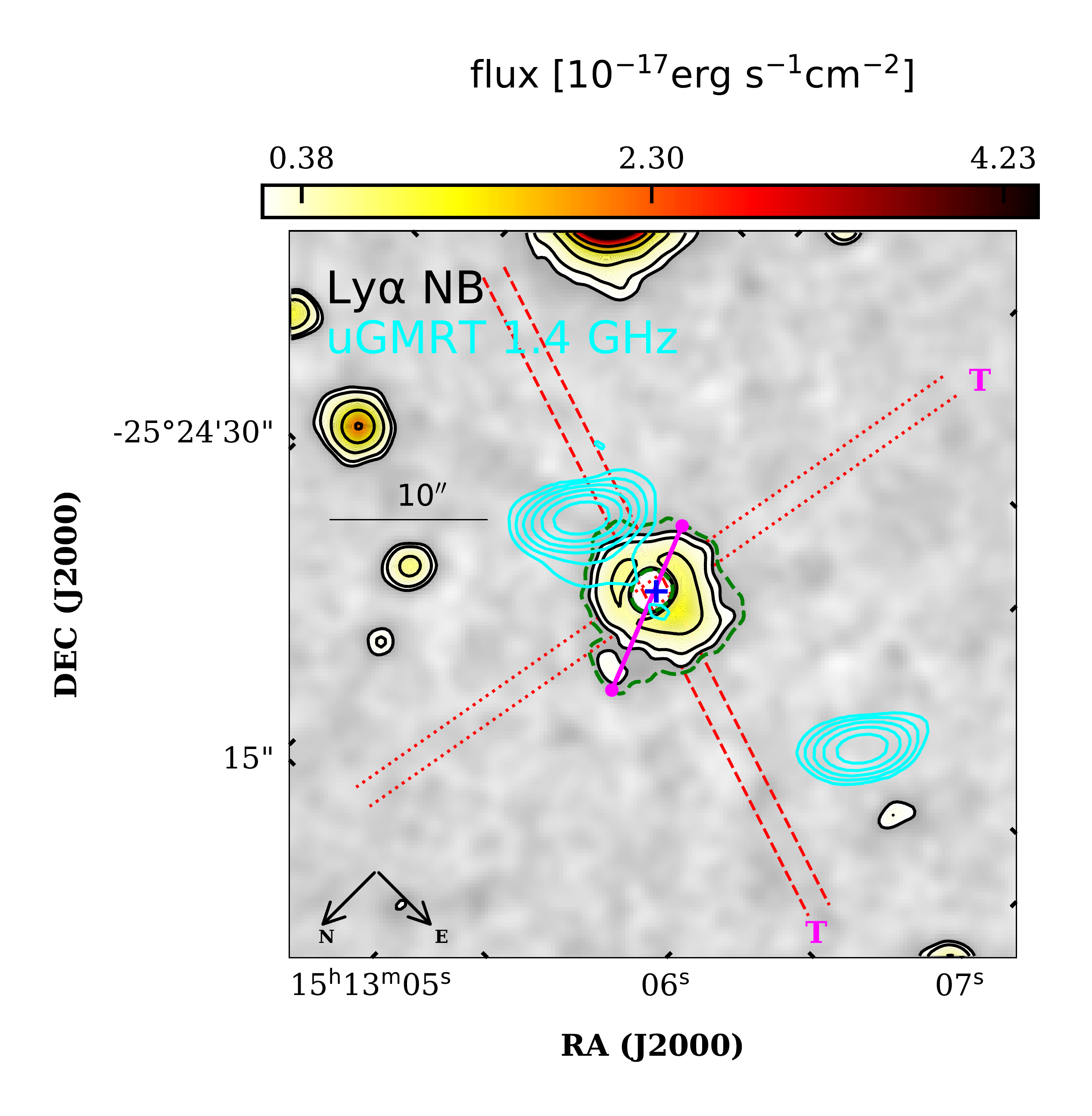}
            
            \caption{{\it Left panel:} \lya\ emission observed in the narrow band image. {\it Right panel:} Narrow band image after subtracting the contribution from the central source. In both the panels the uGMRT Band-5 radio contours (cyan color) are overlaid on top and drawn at the same levels as in Fig. \ref{fig_ps1}. The dashed green contour, corresponds to a 3\sig\ level ($1.06\times 10^{-18}$ \ergscm) and black contours are drawn at (1.6, 2.5, 5.5, 11 and 20 )$\times 10^{-18}$ \ergscm. The flux levels quoted are values per pixel. The position of the WISE source is marked by a blue plus symbol. North and East directions are shown at lower left. The images are smoothed by a Gaussian kernal of FWHM=1.2\arc. The pink line joining two points gives the maximum extent of the \lya\ blob at the 3\sig\ flux level. The dashed red line indicates PA= 72\degree, and the dotted red line indicates PA= 350\degree\ (also see Fig. \ref{fig_ps1}).
            }
         \label{fig_nbcontour}   
        \end{figure*}

The seeing FWHM measured using point sources in the final image, estimated using a two dimensional Gaussian fit, is \apr\ 1.9\arc. The narrow band image is oversampled (0.1267\arc per pixel resolution) compared to the long-slit observations (0.2534\arc per pixel resolution). To account for this oversampling effect, we binned the combined narrow band image by 2\cross2 pixels, resulting in a common pixel scale of 0.2534\arc. We also smoothed the narrow band image using a Gaussian kernal of FWHM=1.2\arc. We used this resampled and smoothed image to quantify the extent of \lya\ emission. In Fig.~\ref{fig_nbcontour}, we have shown the final combined narrow band image of M1513-2524 after smoothing. The uGMRT  Band-5 (1.4 GHz) contours are plotted in cyan. For the \lya\ emission, the 3\sig\ contour is plotted as a green dashed line, and black contours are at 5, 8, 18, 40 and 70 \sig\ levels. The WISE position of the source is marked by a blue plus. The distance  measured between the flux weighted centroid of the \lya\ emission and the WISE location of quasar is  0.26\arc\, which is \til\ 2 kpc at $z=3.13$. Thus, within the astrometric accuracy, the \lya\ peak is coincident with the infrared source.

The flux calibration of the narrow band image was performed using the spectra of the two reference stars (S$_1$ and S$_2$ observed with long-slit observations, shown in Fig.~\ref{fig_ps1}) and their broad band photometry obtained from PS1. Given the wavelength coverage of our long-slit observations, we chose to use only r-band magnitude for this purpose. After comparing the PS1 r-band flux with the flux measured from the long-slit spectra within the wavelength range of the r-band, we obtained slit-loss factors for individual exposures at each PA. First we combined the slit-loss corrected exposures corresponding to each PA. We then used combined slit-loss corrected star spectra of the two PAs to find the total flux (f$^*_\mathrm{tot}$) within the wavelength range covered by the narrow band observations. To find the observed total counts of the stars from the narrow band image, we estimated aperture sizes from radial surface brightness profiles. For each reference star, the aperture radius was fixed as the radius where surface brightness reaches 2\sig\ relative to the background. A simple scaling between the total flux of the star, f$^*_\mathrm{tot}$, and the total counts obtained from the narrow band image of the same star, gives flux per unit count, f$\mathrm{_c = 3.86\times 10^{-19}\ and\ 4.50 \times 10^{-19}} $ \ergscm\ for PA=72\degree\ and 350\degree, respectively. We use the average of these two values, f$\mathrm{_c = 4.18\times 10^{-19}}$ \ergscm,  to flux calibrate the narrow band image. The counts to flux conversion factors (f$\mathrm{_c}$) estimated from the two PAs are consistent with each other within 15\%. To estimate the surface brightness limit (SB$_{lim}$) reached in our combined, resampled and smoothed narrow band image, we put boxes of size 2.5\arc x 2.5\arc\ in background locations and estimate the SB in each box. The mean and rms of these SB values provides SB$_{lim}$. The 2\sig\ SB$_{lim}$ derived using this method is $\mathrm{8.63\times 10^{-18}}$ \ergscma.

\vskip 1cm
%
%
\begin{table*}
 \centering
 \begin{minipage}{155mm}
{\small
\caption{Measurements from the long-slit observations}
\label{tab_longslit}
\begin{tabular}{@{}clc ccrrccc@{}} 
\hline 
 \multicolumn{1}{l}{PA}  
& \multicolumn{1}{c}{f$_{Ly\alpha}$} 
& \multicolumn{1}{c}{L$_{Ly\alpha}$}
& \multicolumn{1}{c}{size$^a$}
& \multicolumn{1}{c}{flux$_{lim}$ }
& \multicolumn{1}{c}{$\mathrm{FWHM_{Ly\alpha}^b}$}
& \multicolumn{1}{c}{$\mathrm{FWHM_{CIV}^b}$}
& \multicolumn{1}{c}{$\mathrm{FWHM_{HeII}^b}$}
\\

&\multicolumn{1}{c}{($\mathrm{10^{-15}\ erg\ s^{-1} cm^{-2}}$)}
&\multicolumn{1}{c}{($\mathrm{10^{44}\ erg\ s^{-1}}$)}
&\multicolumn{1}{c}{(kpc)}
&\multicolumn{1}{c}{($\mathrm{10^{-18}\ erg\ s^{-1} cm^{-2}}$)}
&\multicolumn{1}{c}{(\kms)}
&\multicolumn{1}{c}{(\kms)}
&\multicolumn{1}{c}{(\kms)}
\\
\hline 
72           & 4.40$\pm0.04$ & 3.95$\pm0.05$    &75 & 1.5$ $ & 1269$\pm17$  & 1363$\pm94$ & 1027$\pm87$\\
350          & 4.80$\pm 0.05$ & 4.30$\pm0.04$    &58  & 1.7$ $ & 1362$\pm16$  &1380$\pm59$ & 962$\pm60$\\
\hline   

\end{tabular}                              
}
\\ 
$^a$ Defined by emission detected at a$>$3\sig\ level. \\
$^b$ FWHM from the fitted 1D Gaussian. \\
\end{minipage}
\end{table*}

\section{Results}
\label{sec_results}
In this section we discuss our findings based on the optical and radio observations presented in previous sections.
\subsection{ Redshift of M1513-2524}
\label{sub_z}

\begin{figure}
 \centering
  \includegraphics[viewport=30 30 1000 700,width=10cm, clip=true]{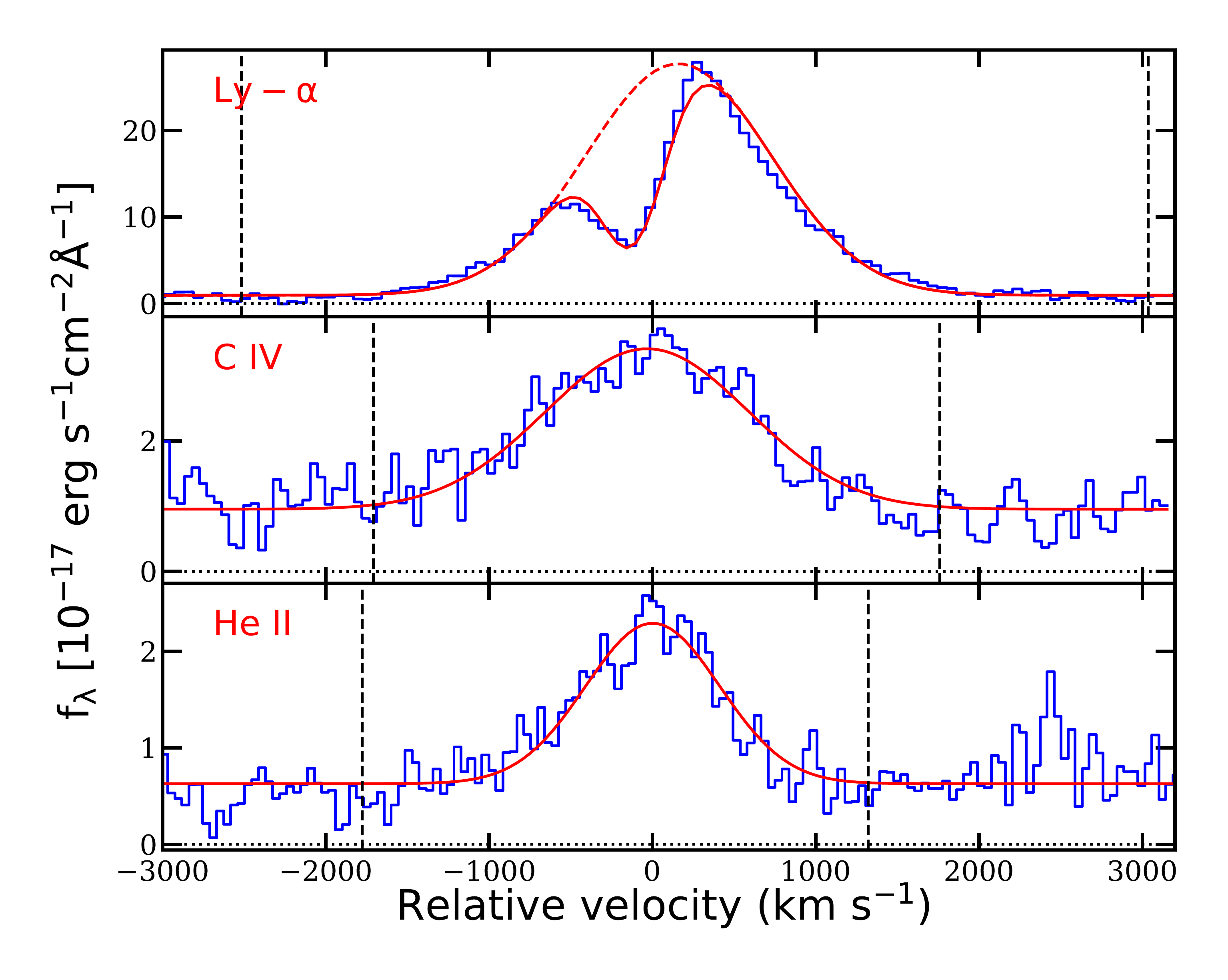}
  
   \caption{Emission line profiles of \lya, \civ\ and \heii\ (blue curves), fitted using 1D Gaussian and Voigt function for \lya\ and 1D Gaussians for \civ\ and \heii. The velocity scale is set with respect to \zem=3.1320 obtained using the \heii\ line. The dotted horizontal line is plotted at zero flux level. Two vertical dashed lines indicate the region over which the line flux is integrated.}
         \label{fig_emission_fit}   
\end{figure}

We clearly detect \lya,  \civ\ and \heii\ emission lines in our long-slit spectrum. N~{\sc v} is also detected at the expected location at a 7\sig\ level (see Fig. \ref{fig_1dspectrum}). The continuum is weak and consistent with the non-detection of the source in the optical broad band photometric images from PS1. The non-resonant \heii\ emission line was used to determine the systemic redshift of the AGN. The redshift was measured using combined 1D spectrum of both PAs, and the \heii\ emission line was modelled using a single Gaussian component, yielding $z_{em}=3.1320\pm0.0003$ (Fig.~\ref{fig_emission_fit}). A recent study by \cite{shen2016} reports that the redshift measured based on the \heii\ line (after taking into account luminosity dependent effects) has a mean velocity shift of 8 \kms\ with an intrinsic uncertainty of 242 \kms\ with respect to the systemic redshift.

\subsection{Emission line ratios}
\label{sub_size}

\begin{figure}
    \centering
    \includegraphics[angle=0,width=0.5\textwidth]{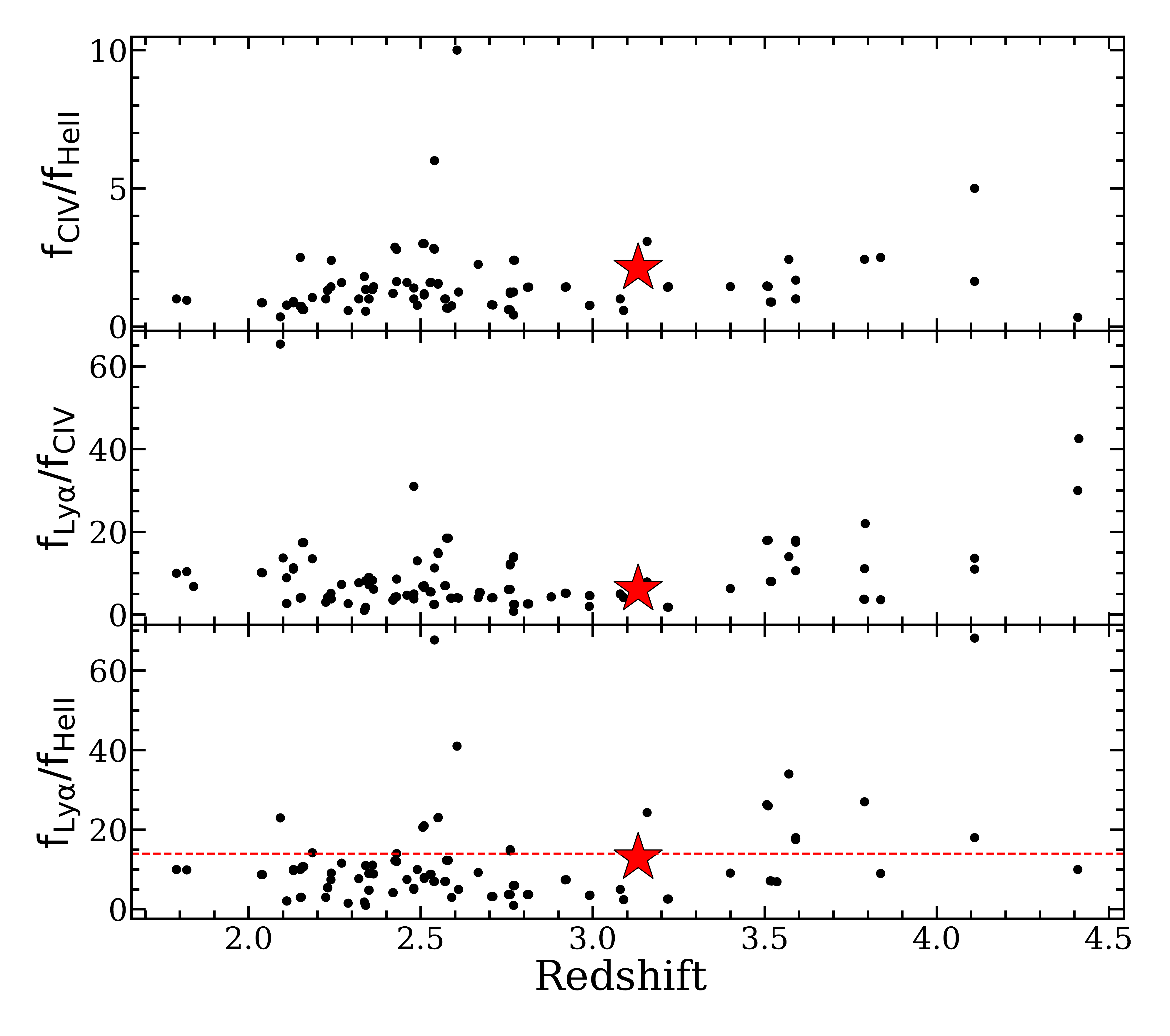}
    \caption{Different emission line flux ratios as a function of $z$ for high-$z$ radio sources from the literature. The location of M1513-2524 is shown with a star symbol in each panel. The dashed red line in the bottom panel is the \flya/\fheii\ cut-off for ``\lya\ excess'' objects (with this ratio $\geq 14$) by definition of VM07. 
    }
    \label{fig:compare1}
\end{figure}

To compute the emission line fluxes and ratios, we used the combined 1D spectrum corrected for slit-loss. The observed \lya\ flux is \flya\ = $(4.93\pm0.04)\times10^{-15}$ \ergscm, and the corresponding luminosity is \llya\ = $(4.42\pm0.04)\times10^{44}$\ergs. For the \civ\ and \heii\ lines, we estimated \fciv\ =$ (8.05\pm0.22)\times10^{-16} $ \ergscm and \fheii\ =$ (3.81\pm0.16)\times10^{-16} $ \ergscm. For each line, the wavelength region over which the fluxes were integrated is marked by vertical dashed lines in Fig. \ref{fig_emission_fit}. The fluxes given above were estimated after subtracting the constant continuum from the line profiles. The \lya\ velocity dispersion (FWHM$_\mathrm{Ly\alpha}$), estimated from the unabsorbed Gaussian component (see Sec. \ref{sub_absorption} for details on fitting) is $1331\pm16$ \kms. The velocity dispersions for \civ\ and \heii\ were estimated after fitting the observed line profiles with single Gaussian components. The estimated velocity dispersions are $1470\pm69$ \kms\ and $968\pm53$ \kms\ for \civ\ and \heii, respectively. The FWHMs quoted here are not corrected for instrumental broadening, which would entail corrections of typically less than 5\%. Note that the FWHM values estimated from the combined spectra obtained with both PAs are slightly different from the values obtained from spectra along individual PAs, given in Table \ref{tab_longslit}, where we have summarized measurements from long-slit observations at both PAs.

Having obtained the line fluxes, we compare the line ratios to understand the nature of the source. The estimated line ratios for \flya/\fciv, \flya/\fheii\ and \fciv/\fheii\ are \til\ 6.12$\pm$0.17, 12.93$\pm$0.57 and 2.11$\pm$0.11 respectively. The ratio of \fnv/\fciv\ and \fnv/\fheii\ are \til\ 0.25\plm0.04 and 0.53\plm0.08, respectively. We point out that, while the line ratios given above are estimated from the slit-loss corrected spectrum, the ratios as expected remain consistent with the values obtained from the spectrum without slit-loss correction. Therefore, the ratios can be directly compared with a literature sample.

In Fig.~\ref{fig:compare1}, we compare our measured ratios with those for a sample of radio sources from the literature. \cite{villar2007a} in particular studied the line ratios and origin of \lya\ emission in a sample of radio galaxies at 1.8$\leq z\leq$4.41. They classified their sample into two catogories, ``\lya\ excess'' objects and ``Non-\lya\ excess'' objects, based on the ratios of \flya/\fciv, \flya/\fheii\ and \fciv/\fheii. In the bottom panel, we plot the \flya/\fheii\ ratio as a function of $z$. From this figure \citep[see also Table 2 of] [VM07 from now on]{villar2007a}, it is clear that the value of this ratio observed for M1513-2524 is lower than the median of measured values for $z\geq3$ sources (median \flya/\fheii = 17.50) and consistent with (or marginally higher than) the values seen in $2<z<3$ radio galaxies. The ratio of \flya/\fheii\ in M1513-2524 is roughly at the expected cut-off for ``\lya\ excess'' objects defined by VM07 (\flya/\fheii\ $\geq 14$, shown as the dashed line in the bottom panel of Fig.~\ref{fig:compare1}). Note VM07 have found about 54\% of the $z>3$ targets in their sample to have the ratio \flya/\fheii $\geq 14$. The large value of this ratio was interpreted as a signature of excess star formation in these objects. Consistent with its ``Non-\lya\ excess" line ratios, the largest angular size (LAS) of the associated radio source in M1513-2524 is much larger than the extent of radio emission associated  with ``\lya\ excess'' objects in VM07. From Fig.~\ref{fig:compare1}, we can see that the flux ratio \flya/\fciv\ is consistent with the median value of the ``non-\lya\ excess" objects (median \flya/\fciv\ = 5.45) (see also Table 3 of VM07). The measured \fciv/\fheii\ is slightly higher than (i.e., within $1.6\sigma$ level) the median values seen for ``Non-\lya\ excess" objects (median \fciv/\fheii\ = 1.0). It is widely believed that photoionization by a central AGN is the cause of \lya\ emission in ``Non-\lya\ excess" objects. Thus, we conclude that the line ratios found in the case of M1513-2524 are consistent with photoionization due to a central AGN.

\subsection{Radio properties}
\label{sub_radio}

In the Band-5 image, at 1.3\arc\ from the WISE position, which is well within the WISE positional uncertainty, a 4$\sigma$ feature (see Fig. \ref{fig_nbcontour}) is detected in the radio continuum. It has a peak flux density of 1.7 $\mathrm{mJy\ beam^{-1}}$.  This feature could be emission from the radio core/jet (i.e., AGN associated with the WISE detection). Deeper and multi-frequency radio interferometric images with higher spatial resolution are required to establish its origin. Our Band-3 image doesn't have adequate spatial resolution for this purpose. Based on the Band-5 image, the largest angular size (LAS), which we define as the separation between the peaks of the eastern and western lobes, is 23.7\arc. The western lobe is closer to the WISE position (i.e., at a linear separation of \til 52 kpc) compared to the eastern lobe (i.e., at a linear separation of  \til 131 kpc). This is the largest radio source in our sample of 25 radio loud AGNs at $z>$2.7. The LAS at the source redshift corresponds to a physical size of 184 kpc for the cosmology assumed here. The position angle measured from North to East is \til\ 96\degree. 

The total flux densities of the eastern and western lobes at 1360 MHz are 48.3 mJy and 117.8 mJy, respectively. The same flux density measured at 420 MHz are 242.1 mJy and 819.4 mJy. Based on these measurements, the eastern and western lobes have spectral indices of $\alpha= -$1.37 and $-$1.65 respectively (for $\mathrm{S_\nu \propto \nu^\alpha}$). These measurements are also summarised in Table~\ref{tab_radio}.

In the lower spatial resolution NRAO VLA Sky Survey (NVSS) image at 1.4 GHz, the total flux density associated with the source M1513-2524 is 217.6 mJy (assuming no variability). This measurement implies that ~25$\%$ of the total radio flux is diffuse and was resolved out in our Band-5 uGMRT image, due to insufficient short baseline coverage and thus lower sensitivity to large spatial scales.

We now use the above measured spectral indices to estimate the flux densities at 150 MHz for both the eastern and western lobes. The estimated flux densities at 150 MHz are \til\ 995 and 4485 mJy for the eastern and western lobes, respectively. To estimate the total radio flux density at 150 MHz associated with M1513-2524, we simply add the two values. This amounts to a total flux density of \til\ 5480 mJy, associated with M1513-2524. However, the total flux density provided by the TIFR GMRT Sky Survey (TGSS) at 150 MHz is only 3632.0 mJy. Further, using the same spectral indices for the eastern and western lobes, we estimate the total flux densities at 365 MHz and 408 MHz to be 1325 mJy and 1110 mJy respectively, whereas the measured flux densities at these frequencies in the literature are ~1220 mJy (see Tabel \ref{tab_photometry}). Within the uncertainties of absolute flux density calibration, these estimates are consistent with our Band-3 measurement. The TGSS flux density measurement then suggests a spectral turnover at $< 300$ MHz in the observed frame (990 MHz in the rest frame).

\setlength{\tabcolsep}{2pt}
\begin{table}
\begin{center}
\caption{Properties of the two radio lobes of M1513-2524 based on the uGMRT observations.
\label{tab_radio}}
\begin{tabular}{lccc}
\hline \hline
\smallskip
\smallskip

Radio component&S$\mathrm{_{1360MHz}}$& S$\mathrm{_{420MHz}}$&$\mathrm{\alpha^{420}_{1360}}$ \\
      &  ($\mathrm{m Jy})$      &  ($\mathrm{mJy})$ &\\
\hline 
Eastern lobe           &  48.3\plm4.8 & 242.1\plm24.2    &-1.37\plm0.12   \\
Western lobe          &  117.8\plm11.8 & 819.4\plm81.9    &-1.65\plm0.12   \\
\hline 
\end{tabular}
\end{center}
\textsc{      } \\
Note: The errors on the flux densities quoted here are dominated by the corrections due to variations in the system temperature.  
We adopt a conservative 10$\%$ uncertainty in the flux densities to account for errors in the flux scale and for the contribution from the thermal noise. 

\end{table}

\begin{table}
\begin{center}
\caption{Flux densities of M1513-2524 from the literature.
\label{tab_photometry}}
\begin{tabular}{c c c}
\hline 
\hline
Survey/Band &  Flux \\
      &  (Jy) \\
\hline 
NVSS 1.4~GHz  & 0.218\plm0.008  \\
 408~MHz \citep{Mills1981}  & 1.22\plm0.08\\
365~MHz \citep{Douglas1996} & 1.22\plm0.04\\
TGSS 150~MHz  & 3.632\plm0.36  \\
\hline 
\end{tabular}
\end{center}
\textsc{} 
 \end{table}

\subsection{Ly$\alpha$ luminosity and excess LAEs}
\begin{table*}
\caption{List of candidate LAEs detected in the narrow band image above 3.0\sig.}
\begin{tabular}{cccccccc}
\hline
\hline
RA&	DEC&	flux per unit wavelength &\llya\ $^a$	&significance$ ~ ~ ~$&	 $m_{r}^\mathrm{{est}}$ &  $m_{lim}^\mathrm{{PS1_r}}$  \\
(HH:MM:SS)  & (DD:MM:SS) &	($\mathrm{10^{-18}}$\fcgs) & ($\mathrm{10^{44}}$\ergs)& (\sig) & &\\    
\hline
15:12:53.6068$ ~ ~ ~$& -25:27:27.021& 1.1\plm0.2 &0.10\plm0.2       &  8.0$ ~ ~ ~$&$ ~ ~ ~$	$ $23.56\plm0.18   &$ ~ ~ ~$	$ $23.33  \\ 
15:13:03.4925$ ~ ~ ~$& -25:25:43.413& 0.5\plm0.2 &0.05\plm0.02       &  3.5$ ~ ~ ~$&$ ~ ~ ~$	$ $24.37\plm0.17   &$ ~ ~ ~$	$ $23.49  \\ 
15:13:01.5284$ ~ ~ ~$& -25:25:20.703& 1.0\plm0.2 &0.09\plm0.02       &  6.7$ ~ ~ ~$&$ ~ ~ ~$	$ $23.69\plm0.20   &$ ~ ~ ~$	$ $23.64  \\ 
15:13:00.5453$ ~ ~ ~$& -25:25:09.402& 1.1\plm0.2 &0.11\plm0.02       &  9.3$ ~ ~ ~$&$ ~ ~ ~$	$ $23.49\plm0.20   &$ ~ ~ ~$	$ $23.64  \\ 
15:13:03.6433$ ~ ~ ~$& -25:26:01.233& 1.3\plm0.2 &0.13\plm0.02       &  10.0$ ~ ~ ~$&$ ~ ~ ~$	$ $23.34\plm0.14   &$ ~ ~ ~$	$ $23.32  \\ 
15:12:59.0238$ ~ ~ ~$& -25:24:07.786& 0.6\plm0.2 &0.05\plm0.02       &  4.7$ ~ ~ ~$&$ ~ ~ ~$	$ $24.30\plm0.30   &$ ~ ~ ~$	$ $23.62  \\ 
15:12:58.3840$ ~ ~ ~$& -25:26:47.314& 1.5\plm0.2 &0.14\plm0.02       &  10$ ~ ~ ~$&$ ~ ~ ~$	$ $23.21\plm0.16   &$ ~ ~ ~$	$ $23.40  \\ 
\hline

\end{tabular}\\
\begin{center}
$^a$\lya\ luminosities  estimated from the narrow band image, assuming fluxes to be that of \lya.\\
\end{center}
\label{tab_lya_objs}
\end{table*}
\begin{figure*}
            \begin{center}
            {\includegraphics[viewport=50 180 850 550,width=20cm,height=10cm,
clip=true,angle=0,clip=true]{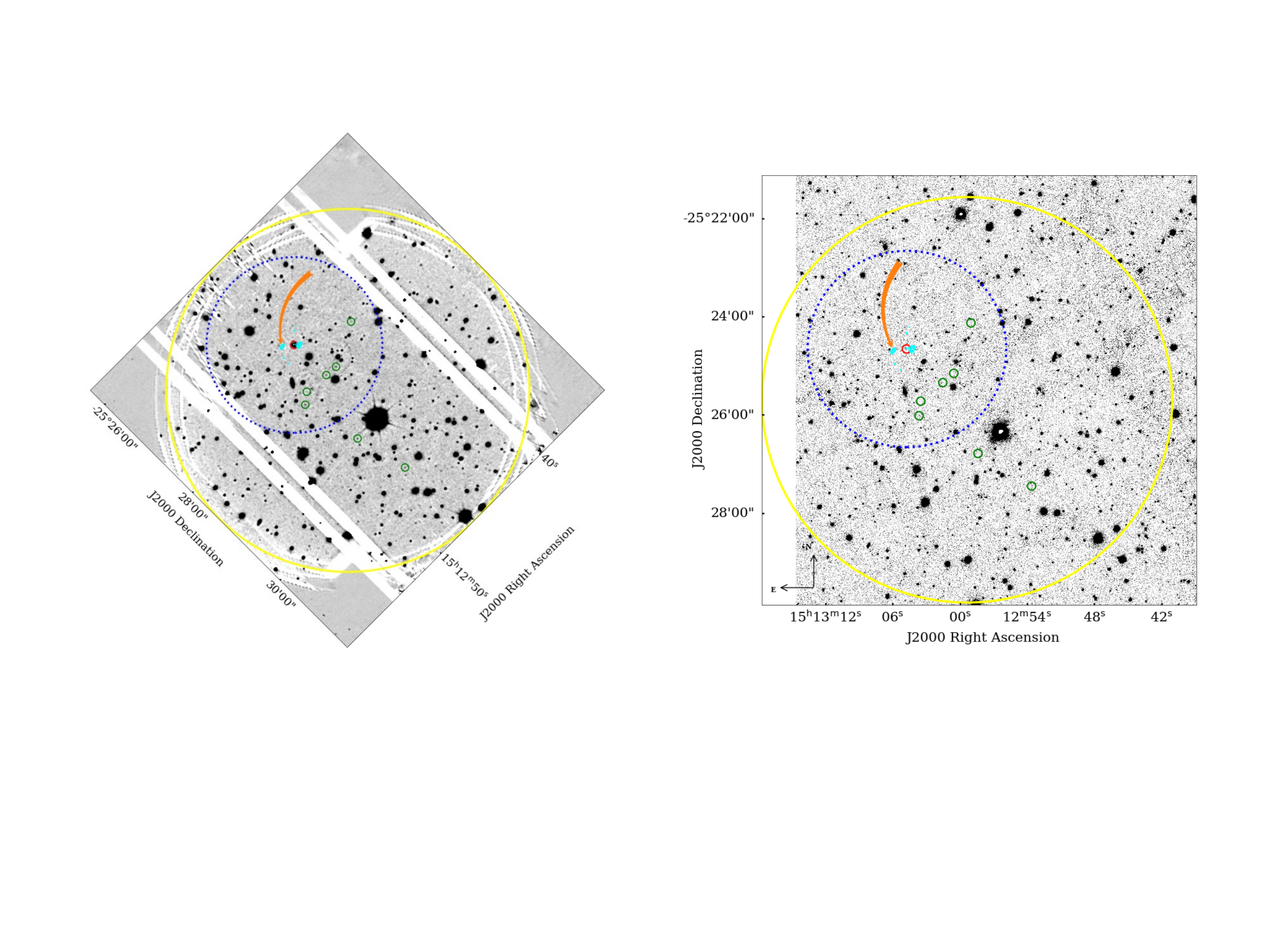}} 
            \end{center}
            \caption{\emph{Left panel:} Combined and smoothed narrow band image of M1513-2524. The image has been rotated so that North is up and East is to the left. \emph{Right panel:} PS1 i-band image. In both the panels, the FoV of our narrow band image is shown by the largest circle (yellow). The 2$^\prime$ region centred on M1513-2524 is shown by dotted the blue circle. The candidate LAEs detected above 3 \sig\ in this FoV are marked by green circles. The red circle is for M1513-2524. The coordinates and magnitudes of these LAEs are given in Table \ref{tab_lya_objs}. The cyan contours (pointed to with the orange arrows) are for the uGMRT Band-5 radio emission, same as shown in Fig. \ref{fig_ps1}.}
            
         \label{fig_nbLAE}   
        \end{figure*}

The total \lya\ flux estimated from the flux calibrated narrow band image within an aperture where the SB reaches 2\sig\ relative to the background is $(7.58\pm0.09) \times 10^{-15}$ \ergscm, and the total \lya\ luminosity L$_{Ly\alpha}$ is $(6.80\pm0.08) \times 10^{44}$ \ergs. The \lya\ flux and luminosity estimated from the PSF subtracted image (details are given in section~\ref{sub_extent}), excluding a central region of radius 1.8\arc, are $(3.11\pm0.08) \times 10^{-15}$\ergscm\ and $(2.79\pm0.07) \times 10^{44}$ \ergs, respectively. Note that the \lya\ emission is at the edge of the narrow band filter response (see Fig. \ref{fig_1dspectrum}). Thus, the actual \lya\ luminosity could still be higher than the above value. In VM07, there are 11 sources which are considered to be ``\lya-excess" objects. We obtained the luminosities for these sources from the quoted fluxes using our cosmological parameters and the median \lya\ luminosity obtained is \til\ $2.09\times10^{44}$\ergs. \emph{ Interestingly, the measured \lya\ luminosity of M1513-2524 is higher than the median \lya\ luminosity of the ``\lya-excess" objects.}

We also provide an estimate for the continuum contribution to the total \lya\ flux. We measure the continuum flux from the 1D spectrum, integrating over the FWHM of the narrow band filter using the wavelength range bluewards of the \lya\ line. The estimated continuum flux is \til\ $(7.4\pm0.4) \times 10^{-16}$\ergscm, which is close to 10\% of the \lya\ flux measured from narrow band image. We also estimate an upper limit on the continuum contribution to the narrow band image using the PS1 r-band image. Here, we estimate the 5\sig\ limiting magnitude at the WISE location of M1513-2524 from the r-band image using an aperture of the same size used to estimate the \lya\ flux in the narrow band image. The estimated r-band limiting magnitude is \til 21.88. The flux per unit wavelength corresponding to this magnitude is \til\ $6.06\times 10^{-18}$\fcgs. Assuming this constant flux over the FWHM of the narrow band filter, the estimated total continuum flux is \til\ $\mathrm{6.7\times 10^{-16}}$ \ergscm, i.e., only 9\% of the total \lya\ flux.

We further check for the presence of other ($\geq3$\sig) \lya\ emitters in the FoV of our narrow band image, that were not seen in PS1 images. We have detected seven \lya\ emitting candidates in addition to M1513-2524 within the FoV (i.e $\sim$ 50 arcmin$^{2}$). The locations of these objects are identified with circles in Fig.~\ref{fig_nbLAE}. We have listed these sources in Table \ref{tab_lya_objs}. The first two columns in this table list the coordinates of the \lya\ emitter candidates. In the third, fourth and fifth columns we provide  narrow band fluxes (per unit wavelength), luminosities assuming these are \lya\ emitters and significance levels of detection respectively. We estimated the expected r-band magnitudes of these sources assuming flat continua (over the wavelength range of the r-band) with flux similar to what we measure in our narrow band image. These values are provided in column 5 of Table~\ref{tab_lya_objs}. The 5\sig\ limiting magnitude reached in the combined PS1 r-band image at the source locations are given in the last column of Table~\ref{tab_lya_objs}. We have used 2\arc \cross 2\arc\ box apertures around the \lya\ candidates to find the fluxes in the narrow band image and 5\sig\ limiting magnitude from the PS1 r-band image. It is evident that in five cases where we detect sources at a $\ge5\sigma$ level, if the signal were due to continuum flux, we would have seen these objects at a $\sim 5\sigma$ level in the PS1 r-band image.  

If we assume the measured narrow band flux is due to \lya\ emission, then the inferred luminosities (see Table \ref{tab_lya_objs}) are close to or higher than what is expected for L* values (\til\ $4.1\times 10^{42}$\ergs) measured for \lya\ emitters at $z\sim3$ \citep{Ouchi2008}. It is expected that there are $\sim10^{-3}$ such galaxies per Mpc$^{-3}$. For the total volume sampled by our narrow band observations, we expect only 0.2 \lya\ emitters brighter than L*. It is also clear from the Figure that five out of the seven identified candidates are within 2$^\prime$ of M1513-2324. Thus if the candidates identified here turn out to be the real \lya\ emitters around $z\sim3.13$, it is highly likely that M1513-2324 is part of a large over-dense region of star forming galaxies. This result would be consistent with what has been found for a few powerful radio galaxies at these redshifts (see references given in the introduction section). Deep multiobject spectroscopic observations are important to confirm these candidates as \lya\ emitters, and verify the presence of excess galaxies around M1513-2324.

 \begin{figure*}
																    
            \begin{minipage}{1\textheight}
            \includegraphics[viewport=0 1 3000 1500,width=21cm,height=10cm, clip=true]{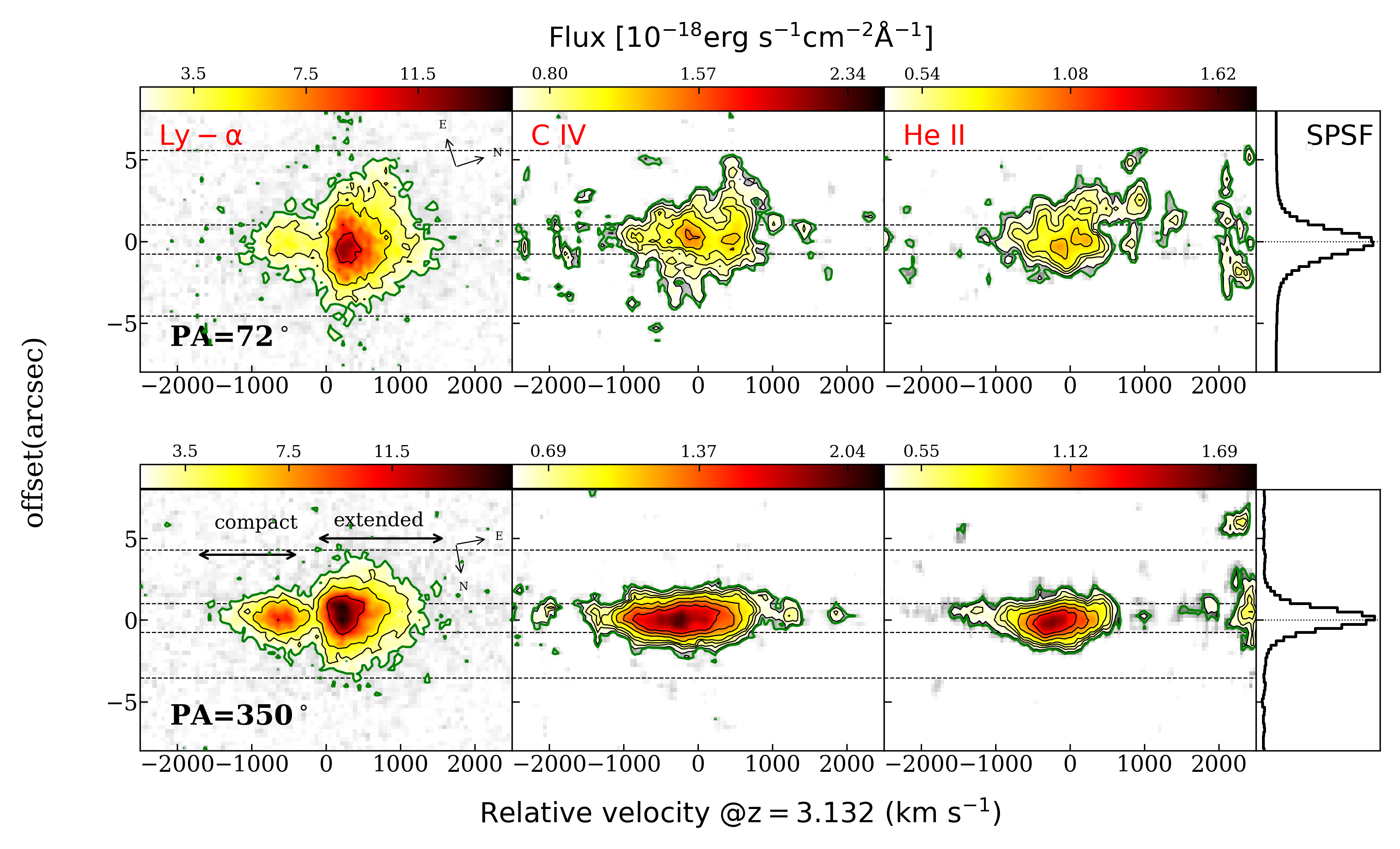}
            
            \end{minipage}
            \caption{2D spectra of M1513-2524 obtained with long-slit oriented along 2 PAs. The upper panel corresponds to PA=72\degree, and the lower to PA=350\degree. In both rows, the first, second and third columns are for \lya\ , \civ\ and \heii\  emission, respectively. The fourth column is the normalized SPSF corresponding to each PA. In the left panel, the thick green contour corresponds to the 3\sig\ level ( $1.59\times 10^{-18}$  and $1.75\times 10^{-18}$ \fcgs\ for PA=72\degree\ and PA=350\degree, respectively). For the \civ\ and \heii\ emission, we have applied boxcar smoothing of 1.27\arc \cross 4.8 \AA\ to improve the signal-to-noise ratio for the plots. The green contours are at the level of 3.5\sig, and subsequent contours are drawn at [4, 5, 6, 7, 9, 11 and 13]\sig\ levels. Note the difference between color bar ranges (top axis of each panel) for \lya\ emission vs. \civ\ and \heii\ emission. The horizontal dashed lines enclose the spatial sections of \lya\ emission used for extracting the spectral profiles shown in Fig. \ref{fig_vel_sections} from the SPSF subtracted spectra. The North and East directions of Fig. \ref{fig_ps1} are shown in left panels.
            }
            
         \label{fig_2dspectrum}   
        \end{figure*}

\vskip 1in
\begin{figure*}
																    
            \begin{minipage}{1\textheight}
            \includegraphics[viewport=0 1 3000 1500,width=21cm,height=10cm, clip=true]{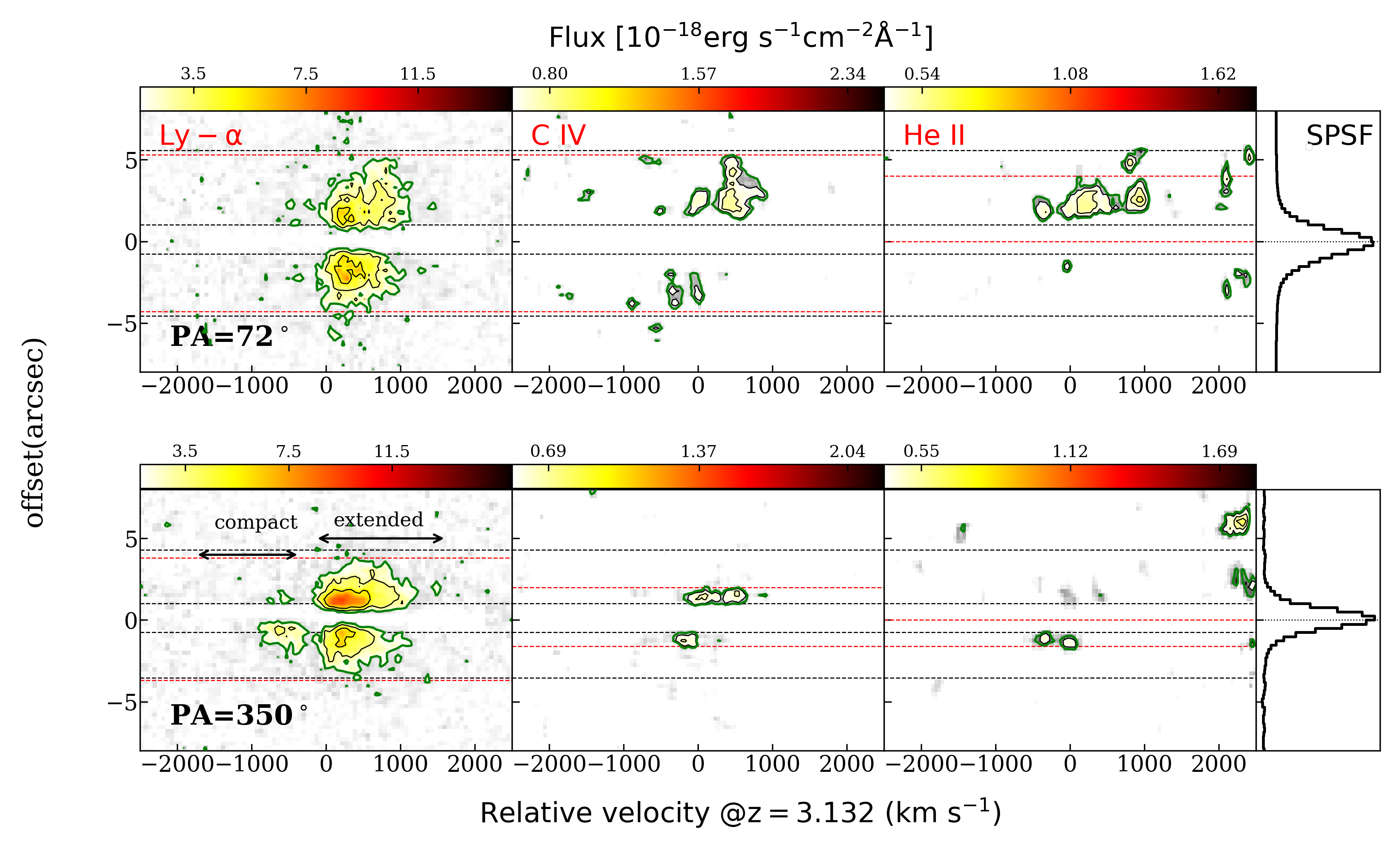}
            \end{minipage}
            \caption{Same as Fig. \ref{fig_2dspectrum}, except that here spectra are plotted after subtracting the SPSF. The contour levels are as shown in Fig. \ref{fig_2dspectrum}. The horizontal red dashed lines indicate the spatial extents of the emission lines discussed in Sec.\ref{sub_extent}. }
        \label{fig_2dspectrum_sub}   
        \end{figure*}

 \subsection{Extent and morphology of line emitting regions}
 \label{sub_extent}

In Fig.~\ref{fig_2dspectrum}, we show the 2D spectra over the wavelength ranges of the three main emission lines observed at the two position angles discussed above. The velocity scales are given with respect to the emission redshift measured using the He~{\sc ii} emission lines (i.e., \zem = 3.1320). On the right hand side of each panel, we also provide the spectral PSF (SPSF). As our observations were carried out in moderate seeing conditions, we perform SPSF subtraction to remove the contribution of emission from the central regions to the extended emission seen at large distances, in order to determine the extent of \lya\ emission. It is clear from this figure that the spatial extent of the \lya\ emission is larger than the SPSF and non-uniform over the wavelength range covered by the emission. The entire \lya\ emitting region can be split into two spatial components, a ``compact'' spatial component [\lya(c), spread over \til\ -367 to -1741 \kms] bluewards of the absorption trough and an ``extended'' spatial component [\lya(e), spread over \til\ -128 to 1603 \kms] redwards of the absorption trough.

To estimate the extent of the \lya\ emission that is unaffected by seeing smearing, we carried out SPSF subtraction from the 2D spectra. For each PA, we construct SPSF from the reference star spectra after collapsing a \til 20\AA\ region close to the \lya\ central wavelength. We then carry out a simple two parameter (i.e., peak amplitude and centriod of the SPSF) fit  using chi-square minimization with respect to the spatial profile of \lya\ emission at every wavelength. We constructed 2D SPSF spectra using the best fit amplitude and centroid. We refine the centroid position as a function of wavelength by linear fit. This was used for the final position where the peak of the SPSF was centred. We then simply subtract these 2D SPSF model spectra from the \lya\ emission. The 2D spectra along PA =72\degree\ and PA=350\degree\ after SPSF subtraction are shown in Fig.~\ref{fig_2dspectrum_sub}.
 
 The extent of diffuse \lya\ emission is estimated from the SPSF subtracted 2D spectra  as the maximum diameter of the 3\sig\ flux contour in the continuum subtracted 2D spectrum (enclosed within two red dashed curve, in Fig. \ref{fig_2dspectrum_sub}).  It is evident from the figure that the \lya(e) component extends up to \til\ 75 kpc at 3\sig\ flux limit of $\mathrm{1.5 \times 10^{-18}}$ \ergscm\  along PA =72\degree\ and \til\ 58 kpc (flux limit of $\mathrm{1.7 \times 10^{-18}}$ \ergscm) along PA =350\degree. We note that the \lya(c) component does not extend beyond the SPSF along PA =72\degree; however, it is asymmetric and extends up to \til\ 19 kpc along PA =350\degree. 

To characterise the morphology and estimate the extent of the full \lya\ emitting region, we use our re-sampled narrow band image (see Fig. \ref{fig_nbcontour}). Again to remove the effects of seeing smearing on the emission measured at large distances, we subtract the PSF from this image as well. To construct the PSF, we have taken a star present in the FoV of the narrow band image. We follow the same empirical method for PSF subtraction used by various studies in the literature  \citep[see, e.g.,][]{Borisova2016,Arrigoni2019}. This method of PSF subtraction is based on simple scaling of the normalised PSF image to match the flux within the central region of \lya\ emission. Usually in the literature studies, the central 1\cross 1 arcsec$^2$ around the quasar is used for finding the rescaling factor. These studies were done under excellent seeing conditions (ranging from 0.59-1.31 arcsec); due to our poor seeing of \til\ 2\arc, we have used the area within the central 2\arc\ radius to find the scaling factor. Before we subtract the rescaled PSF, the centroid of the PSF image is aligned with the flux weighted spatial centroid of the \lya\ region of M1513-2524 above 3\sig. The area chosen is such that there are no other sources within this region. We measure the \lya\ extent from the PSF subtracted image after smoothing it using a Gaussian kernel of FWHM=1.2\arc. The smoothed narrow band images before and after PSF subtraction are shown in the left and right panel of Fig. \ref{fig_nbcontour}, respectively. The pink line joining two points in Fig. \ref{fig_nbcontour} shows the maximum \lya\ extent at 3\sig\ (flux limit= $\mathrm{1.06\times 10^{-18}}$\ergscm), equal to \til\ 90 kpc. Note that the largest \lya\ extent seen in the narrow band image doesn't align with the radio axis and the angle between the radio axis, and the direction where we see the largest \lya\ extent is \til\ 72\degree.

\begin{figure*}
           \centering
            \includegraphics[viewport=10 35 1390 1050,width=16cm, clip=true]{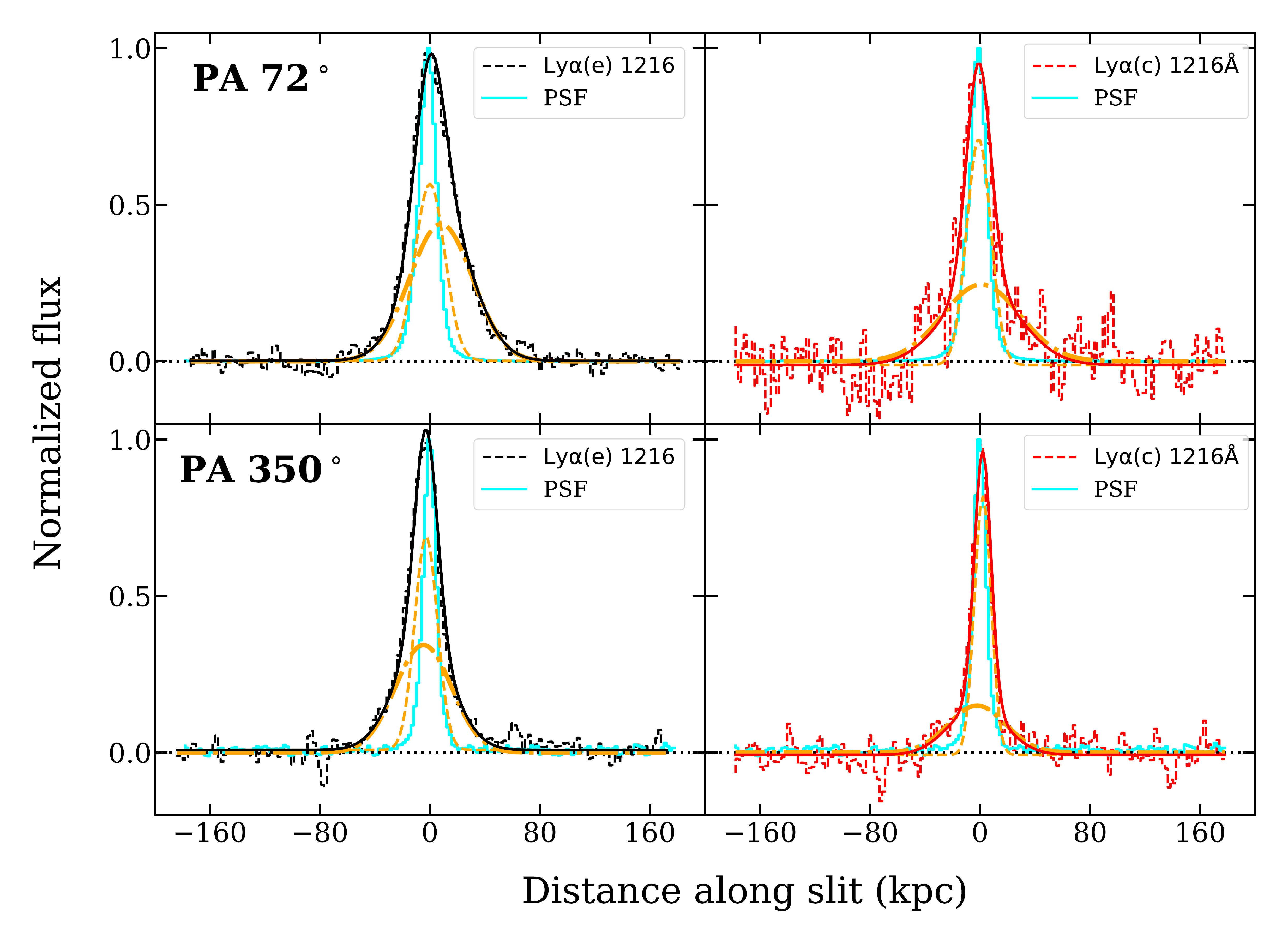}
            \caption{Comparison of spatial profiles of \lya\ with SPSFs extracted from long-slit spectra along PA = 72\degree\ and PA = 350\degree. The profiles are extracted from two regions: a ``compact'' spatial component (marked c, red) and ``extended'' spatial component (marked e, black) (see Fig. \ref{fig_2dspectrum}). Dashed cyan curves correspond to SPSFs from reference stars. The actual spatial profile for each emission line is shown as a dashed and the corresponding overall 2-component Gaussian fit is shown by a solid curve. In each subplot the orange dashed and dash-dotted lines are 2 Gaussian components fitted to the corresponding emission line. The black dotted horizontal line marks the zero flux level. The vertical shift below zero level (especially seen in upper panels) is due to artifacts in the data.
             }
            
            \label{fig_line_comparison1} 
        \end{figure*}

In Fig. \ref{fig_line_comparison1}, we have also shown a direct comparison of the spatial profiles of \lya(c) and \lya(e) before SPSF subtraction, for better visualization of the extent of these two components. It is clear from these figures that the \lya(e) component extends much beyond the SPSF. \lya(c) component also seems to possess an extended component along PA = 350\degree, which is consistent with the result discussed above using 2D SPSF subtracted image. However, \lya(c) at PA =72\degree\ is noisy, and extended \lya\ can not be confirmed in this case.

\begin{figure*}
           \centering
            \includegraphics[viewport=10 1 1390 1050,width=16cm, clip=true]{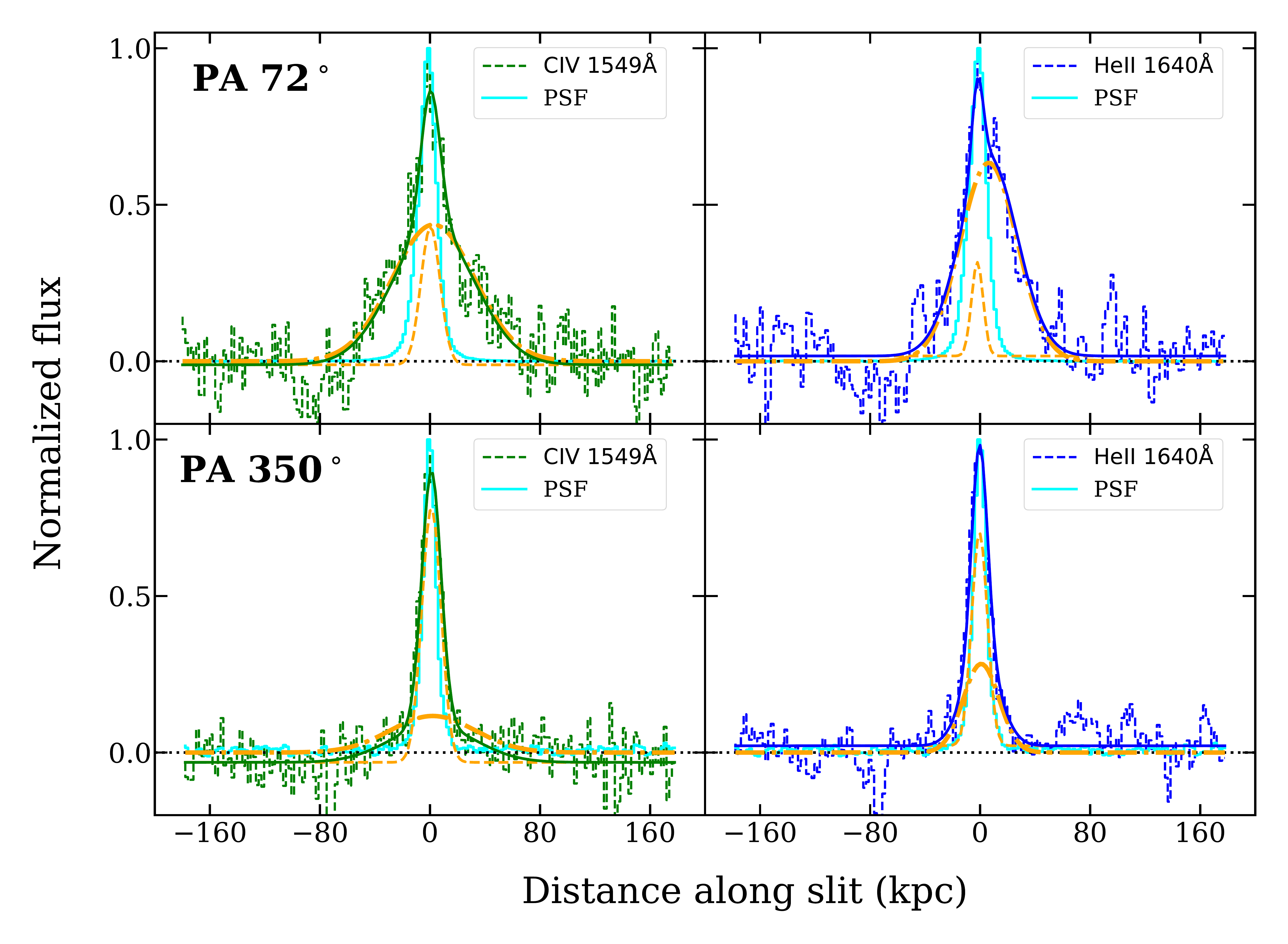}
            
            \caption{Comparison of spatial profiles of \civ\ and \heii\ with SPSFs from two PAs. For \civ\ and \heii\ the full emission regions were used to extract spatial profiles. Dashed green and blue curves correspond to the observed \civ\ and \heii\  lines, respectively.}
            \label{fig_line_comparison2}  
        \end{figure*}

Apart from the detection of the extended \lya\ blob, we investigate if \civ\ and \heii\ lines are also spatially extended, which will be important for understanding the physical conditions prevailing in the extended line emitting region. These lines are also important for constraining the kinematics of the gas, as the \lya\ profile is sensitive to radiative transfer effects. To draw a conclusion on the extension of \civ\ and \heii\ lines, we subtract the SPSF from their 2D spectra as well (see Fig. \ref{fig_2dspectrum_sub}). It is clear  from the the SPSF subtracted images that both \civ\ and \heii\ lines are clearly extended along PA =72\degree.  In both cases, the residual emission found is asymmetric, with more extended emission on one side than the other. However, these two lines do not appear extended along PA =350\degree\ beyond the SPSF. Thus, our data suggest an asymmetric distribution of extended \civ\ emission. In particular, it is interesting to note the direction in which we find the maximum extension (i.e., PA = 72\degree) is closer to the radio axis (i.e., PA = 96\degree). To explore this issue further, in Fig. \ref{fig_line_comparison2}, we have also shown a comparison of the spatial profiles of \civ\ and \heii\ with the SPSF for the two PAs. Based on this analysis, it is clear that both \civ\ and \heii\ emission are significantly extended along PA =72\degree. However, for PA =350\degree, the significance of the detection of extended emission for \civ\ and \heii\ is not high. In Fig. \ref{fig_line_comparison1} and \ref{fig_line_comparison2}, we have also shown a 2-component Gaussian fit to the observed spatial profiles, demonstrating the presence of extended emission at large spatial scales. Thus, based on our long-slit spectroscopy it appears that the \lya, \civ\ and \heii\ emission may not be isotropic in nature, with the largest extent perhaps aligned close to the radio axis. Note in our narrow band image, the \lya\ extent is not aligned with the radio source. Therefore, it will be interesting to acquire IFS observations of \lya\ and \civ\ emitting regions to study the anisotropic distribution of gas excitation and its connection to the radio emission \citep[see for example,][]{mccarthy1993}.

To quantify the 2D morphology of the \lya\ halo, we measure the  asymmetry parameter $\alpha=b/a$ from the unsmoothed narrow band image using the procedure explained in \cite{Arrigoni2019} (here after A19). The asymmetry parameter $\alpha$ is obtained using the second-order moment of the light distribution of the nebula, where $a$ and $b$ are the semi-major and semi-minor axes, respectively. We obtain $\alpha=0.84$ for the region within the 3\sig\ isophote. This value suggests a symmetric distribution as found for majority of high-\emph{z} quasars \citep{Arrigoni2019}.\\

As discussed above, the 2D spectra and the spatial profiles suggest the existence of two components, i.e., a compact core region and an extended \lya\ halo region. The narrow band image is the superposition of these two components. To test whether the \lya\ emission is isotropic even in the central regions, we consider the isophote corresponding to the flux level of $1.42\times10^{-17}$\ergscm (i.e., 10\sig\ level of the background) in the unsubtracted narrow band image. We find $\alpha=0.79$ for this isophote. Interestingly, the semi-major axis of this isophote is aligned close to the radio axis and the radius at 3\sig\ and 10\sig\ isophote are \til\ 4.6\arc\ and 2.1\arc\ respectively. Thus, our data suggest that the \lya\ emission is symmetric in the outer regions and shows a trend of increasing asymmetry as one moves towards the central region i.e., close to the AGN.

\begin{figure*}
	    \includegraphics[viewport=20 35 1400 700,width=18cm, clip=true]{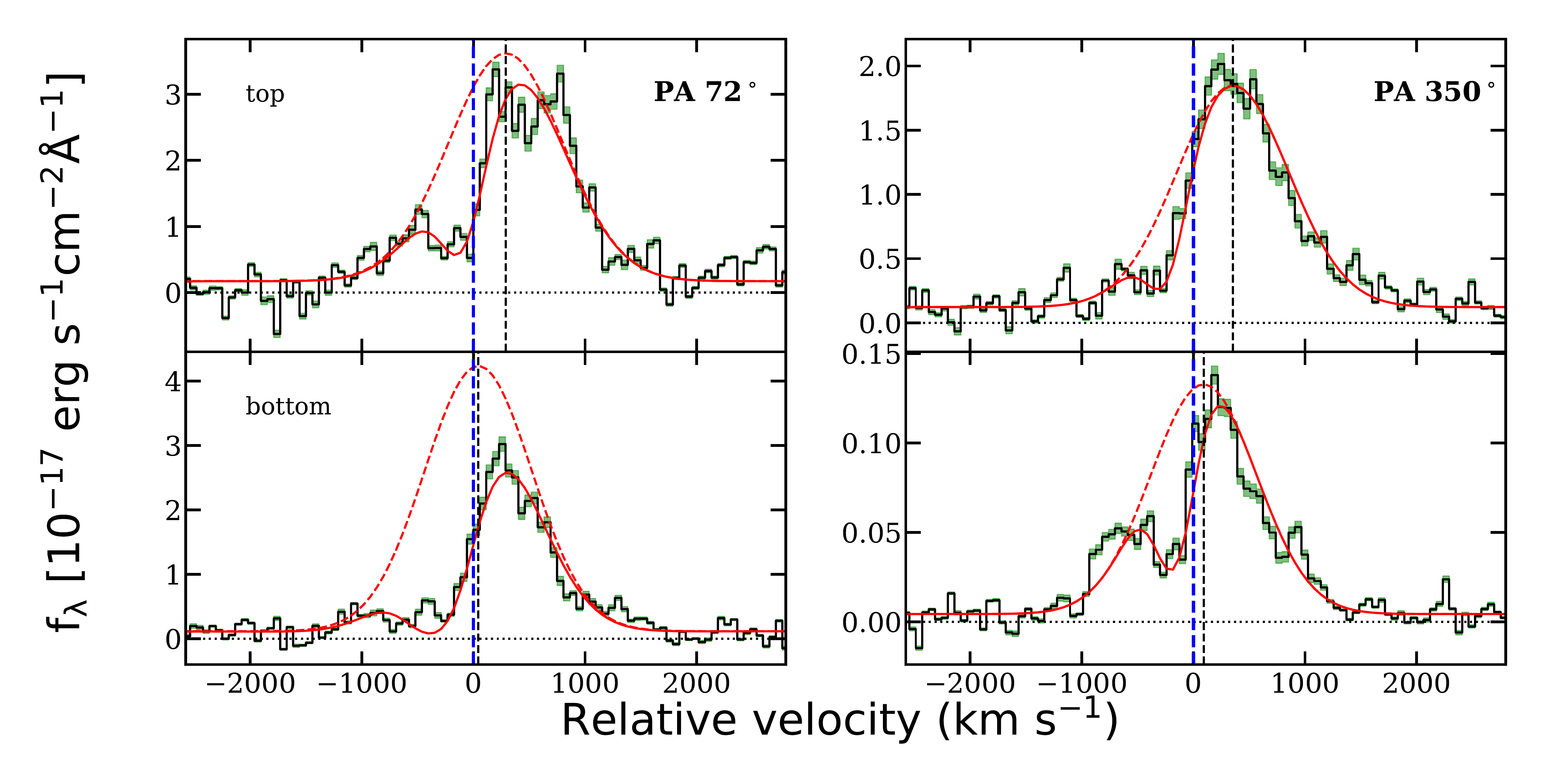}
            \caption{\lya\ emission profiles extracted from the top and bottom spatial regions of the PSF subtracted images for two PAs (see Fig. \ref{fig_2dspectrum_sub}). The observed profiles are fitted with a 1D Gaussian+Voigt function. The black solid line is the observed profile, the red solid line is the absorption model fit and the dashed line is the unabsorbed Gaussian profile. The left and right panels correspond to PA= 72\degree\ and PA=  350\degree, respectively. The gray shaded regions are 1\sig\ error bars. The zero velocity (blue dashed line) is defined with respect to the emission redshift measured using \heii\ line of the combined 1D spectrum of both the PAs. The black dashed lines mark the peak of unabsorbed Gaussian fit. 
            }
            
         \label{fig_vel_sections}   
        \end{figure*}
\subsection{Gas kinematics}
\label{sub_gas}

To get more information on the velocity distribution of the \lya\ emitting gas, we take spatial sections along the slit in the SPSF subtracted \lya\ spectra and measure the center and FWHMs of the fitted Gaussian profiles for these sections. The details of the fit used for each section are described in section \ref{sub_absorption} (see Fig. \ref{fig_vel_sections} ``top'' corresponds to East for PA=72\degree\ and South for PA=350\degree). It seems that the peaks of \lya\ emission on the top are clearly redshifted with respect to the measured redshift of M1513-2524. However the bottom regions are only slightly redshifted.

As the \lya\ emission profile is sensitive to radiative transfer effects in addition to the underlying velocity field, it will be difficult to interpret the velocity field purely based on the \lya\ profile alone. However, it is also clear from the SPSF subtracted spectra that the residual \civ\ emission we see in spectra obtained at PA = 72\degree\ is also consistent with the extended emission at the top being redshifted (see Fig.~\ref{fig_2dspectrum_sub}) with respect to the systemic redshift measured from the \heii\ line. But the residual emission seen in the bottom part is either consistent with the systemic redshift or slightly blueshifted. This result is in line with what we see in the case of the \lya\ profile. When we consider the 2D spectra taken at PA = 350\degree, the residual \civ\ emission in the top is weak. However, we do see a trend similar to what we see for the \lya\ line. Thus, there are signatures in the data indicating that the extended emission to the southeast is redshifted with respect to the systemic redshift. Given the fact that we are seeing blue shifted \lya\ absorption, the flow in the surrounding regions of M1513-2524 may be complex. We note that the FWHM is highest at the center (\til\ 1300 \kms) and decreases away from the center (ranging from 800-1000 \kms). Interpreting this sparsely sampled velocity field in terms of infall, outflow or rotation will be difficult. High spatial resolution ($<$1\arc) IFS data are required to reveal further details on the velocity field and spatial extent of the core and extended spatial components.

%
\vskip 1cm 
\subsection{Connection between radio and Ly$\alpha$ emission }
\label{sub_connection}

As seen above, the alignment of \civ\ and \lya\ with the radio axis is indicated by our data. This agreement could indicate good alignment between the putative cone of ionizing radiation and the radio jets, and/or excitation due to interaction of radio emitting plasma with ambient gas. Broad emission line widths may indicate possible interactions. In this section, to gain more insight, we will compare the properties of M1513-2524 (both optical and radio) with those of radio sources in the literature. For this comparison, we have used results from the literature \citep{roettgering1994,vanojik1997,debreuck2001,jarvis2001,willot2002,bornancini2007,saxena2019}, with a view to understand the effect of radio jets on the ambient medium. High redshift radio galaxies have shown strong dependence on \lya\ size and velocity dispersion with the associated radio sources \citep{vanojik1997}. Typically, small \lya\ halos are seen to be associated with smaller radio sources showing higher velocity dispersion. For low-\emph{z} radio galaxies \citep{mccarthy1991}, brightest regions of \lya\ emission are seen to lie on the side of the radio lobe closest to galaxy nucleus. However, \cite{vanojik1997} found that the brightest regions of \lya\ emission are not always on the side of the radio lobe closest to the nucleus, but in most cases lie closest to the brightest lobe.

From Fig. 6 of \citet{saxena2019} we find that radio galaxies with \lya\ luminosity $\ge 3\times 10^{44}$ erg s$^{-1}$ typically have FWHM velocities in excess of 1000 \kms. Thus, what we find for M1513-2524 is consistent with that of the sample studied by \citet{saxena2019}. As we have shown above, the extended \lya\ emission from M1513-2524 is symmetric (when we consider the 3$\sigma$ contours), and the radio lobes are mostly outside the diffuse \lya\ emission for the SB sensitivity we have achieved in our narrow band images (see Fig. \ref{fig_nbcontour}). However, there are indications that the \lya\ emission in the central region may be asymmetric and probably elongated along the radio axis. It is also seen that \civ\ emission is more extended close to the radio axis. Thus, we do see indications for connections between the diffuse spectral line emission and radio emission. However, better spectroscopic data (in terms of both spatial and spectral resolution) will be useful to quantify these observations at a higher significance level.

\begin{figure}
    \centering
    \includegraphics[viewport=25 35 1600 850,width=12cm, clip=true]{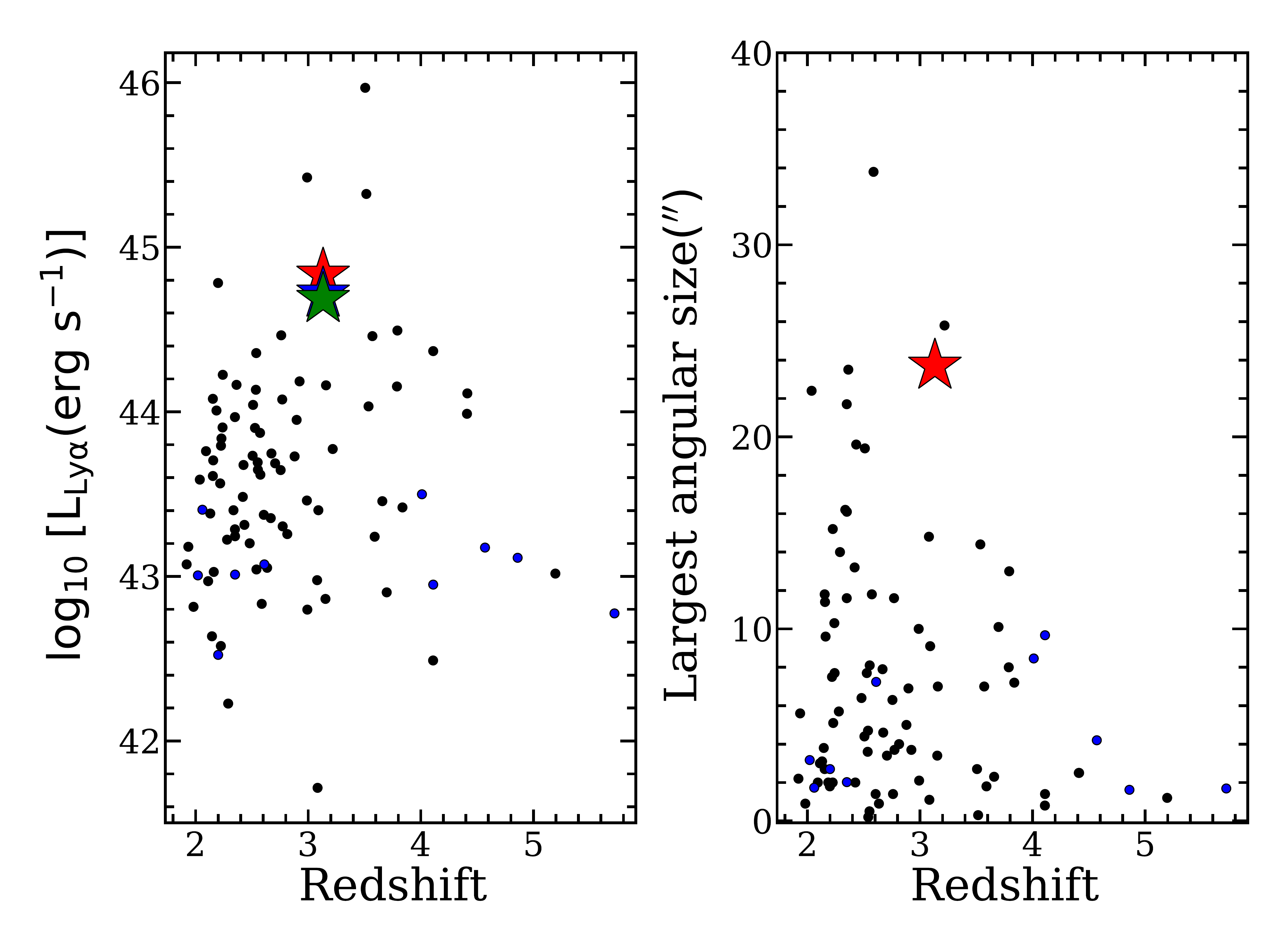}
    
    \caption{\lya\ line luminosity (left) and LAS (right) of the radio emission as a function of $z$ for high-$z$ radio sources from the literature. Location of M1513-2524 is shown with a star symbol. In the left panel, the red star is the \lya\ luminosity estimated from the narrow band image, the blue star is estimated from slitloss corrected combined spectrum of both PAs, and the green star is from the slitloss-uncorrected spectrum.}
    \label{fig:compare}
\end{figure}

\begin{figure*}
            \begin{minipage}{1\textwidth}
            \includegraphics[viewport=5 35 1800 750,width=18cm, clip=true]{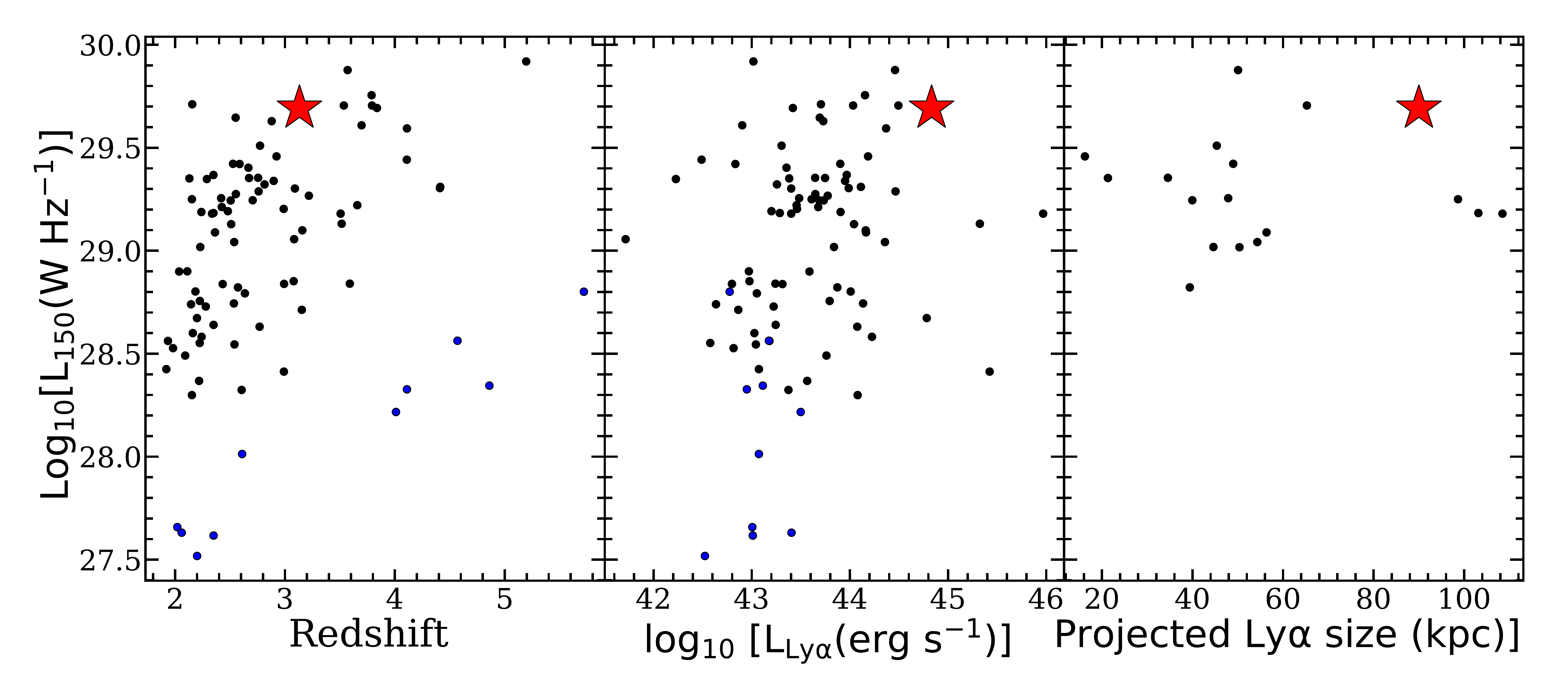}
            \end{minipage}
            \caption{Radio luminosity at 150 MHz is shown as a function of redshift, \lya\ luminosity and projected \lya\ size for M1513-2524 and a sample of faint HzRGs (blue points, \citet{saxena2019}) and normal HzRGs (black points). In the rightmost panel, only data sample from \citet{vanojik1997} is shown, since radio source size information were 
            not readily available for other sources.
            }
     \label{fig_compare4}
       \end{figure*}

In Fig.~\ref{fig:compare} we plot the measured \lya\ luminosity and largest angular scale of the radio source as a function of redshift for known $z>2$ radio sources  with \lya\ measurements in the literature. It is clear from the left panel of this figure that only three (TNJ0121+1320, TNJ0205+2242 and WN 2313+4053) literature sources at $z>2$ have \lya\ flux more than what we measure for M1513-2524. All these sources are less extended in radio emission compared to M1513-2524. In the right panel of Fig.~\ref{fig:compare}, we plot the largest angular sizes of radio sources as a function of $z$. Even in this plot, only two (TN~J1123+3141 and TXS~J2335-0002) literature sources have larger radio structures compared to M1513-2524. However, the \lya\ fluxes measured for these sources are at least a factor 10 smaller than what we measure in M1513-2524. We note that the \lya\ emission of the other high-$z$ radio galaxies was measured mostly using long-slit observations, complicating comparison with a \lya\ luminosity of M1513-2524 estimated from narrow band imaging. Therefore, in the left panel of Fig. \ref{fig:compare}, we have marked the \lya\ luminosity obtained from narrow band image (red star), slitloss-corrected combined 1D spectrum of both PAs (blue star) and combined spectrum before slitloss correction (green star). {\it Thus, M1513-2524 indeed appears to be rare in terms of its radio size and its Ly$\alpha$ luminosity.}

In Fig. \ref{fig_compare4}, we have shown a comparison of \lya\ luminosity and radio luminosity at 150 MHz for M1513-2524 with sample of radio sources from the literature. \citet{jarvis2001} have shown a strong correlation between the \lya\ emission line luminosity and radio luminosity measured at 150-MHz (L$_{150}$) for radio galaxies at $z>1.75$ \citep[see also the figure 8 of][]{saxena2019}. From the left panel of Fig.~\ref{fig_compare4} it is clear that M1513-2524 has a large 150 MHz luminosity compared to most high-z radio sources. In the middle panel, we plot L$_{150}$ vs. L$_{\lya}$. The measured L$_{150}$ for M1513-2524 is 3.08$\times10^{29}$ W Hz$^{-1}$. For this value, the expected \lya\ luminosity from the fit presented in \citet{saxena2019} is \til 3.16$\times10^{43}$ \ergs. The \lya\ luminosity we observe is much higher than this best fit prediction, as also evident from the middle panel. While there are 3 sources in the literature sample with \lya\ luminosity higher than what we find in M1513-2534, all of them have  L$_{150}$ less than what we measure for M1513-2534.

In the right panel of Fig.~\ref{fig_compare4}, we plot the projected \lya\ size vs radio power at 150 MHz. M1513-2524 has one of the largest \lya\ extents among the strong radio sources (where the extents were mostly measured from  long-slit spectroscopy observations). However, recent studies based on radio weak quasars \citep{Borisova2016,Arrigoni2019} have shown \lya\ emission extending beyond 100 kpc. These objects were observed with much better surface brightness sensitivities.  However, a direct comparison of M1513-2524 with these sources is rather difficult, as different sets of data were obtained with different surface brightness sensitivities. If the surface brightness profile of M1513-2524  is not very different from other high-$z$ quasars one will expect the \lya\ extent to be much larger than what we measure. Therefore, obtaining deep IFS spectra of M1513-2524 will be important for ascertaining the true extent of the \lya\ emission.

\subsection{Associated H~{\sc i} absorption}
\label{sub_absorption}
 
Associated \lya\ absorption is clearly evident in the \lya\ emission line profile. However, we do not detect any associated C~{\sc iv} absorption in the profile of the C~{\sc iv} emission line. As we discussed above, the extent of \lya\ emission is not uniform over the wavelength covered by emission (see Fig~\ref{fig_2dspectrum}). It is clear from the figure that \lya\ absorption occurs roughly in the wavelength range that demarcates the ``compact'' and ``extended'' spatial components. This basically limits the spatial extent over which we can probe the \lya\ absorption. Under the assumption that the observed \lya\ emission profile is the result of an unabsorbed Gaussian profile modified by the presence of neutral hydrogen (\hi) along the line of sight, which can be characterised by a Voigt function \citep[see][hereafter VO97]{vanojik1997}, we estimate the properties of the \hi\ absorption. The initial input parameters for the model are amplitude, redshift and width of the 1D Gaussian, and absorption redshift (\zabs), Doppler parameter ($b$) and column density (log $N$(\hi)) for the Voigt profile.

\begin{figure*}
 \centering
  
  \includegraphics[viewport=20 0 2500 1400, width=20cm,clip=true]{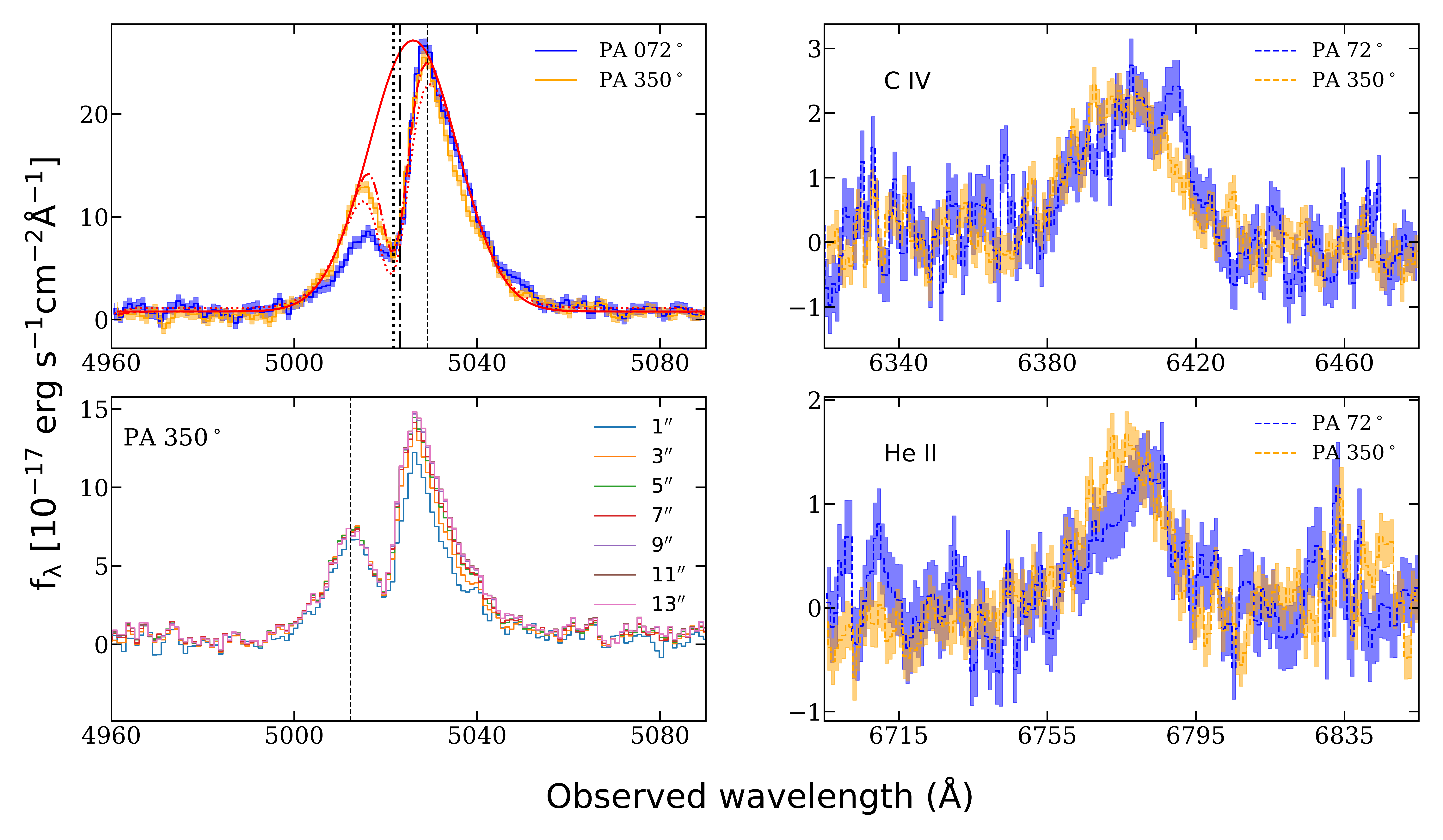}
  
   \caption{\emph{Top left panel:} Comparison of \lya\ velocity profile at 2 PAs is shown after scaling the spectra to match at the peak of narrow \lya\ component (at thin vertical dotted line). The blue and orange solid curves correspond to spectra obtained at PA = 72\degree\ and 350\degree, respectively and the shaded region shows 1\sig\ region. The corresponding fits to the observed profiles are shown in dash-dotted red color (PA = 72\degree) and dotted red curve (PA = 350\degree). The Gaussian emission line profile obtained from our chosen model (Gaussian plus Voigt profiles) at PA = 350\degree\ is shown in red solid curve. For PA = 72\degree\ same Gaussian profile is used with varying the column density, doppler parameter and absorption redshift. The thick vertical dotted line marks the wavelength corresponding to absorption redshift and the dash-dotted line marks the wavelength correspondng to the emission redshift obtained using \heii\ line. \emph {Bottom left:} The \lya\ profiles obtained from spectrum at PA = 350\degree\ using different spatial apertures with varying width after scaling the fluxes to match at the location of the dotted vertical line are shown. \emph{Right panel:} \civ (top) and \heii (bottom) profiles obtained along two PAs are compared.} 
 
\label{fig_PAs_vel_prof}   
\end{figure*}

In upper left panel in Fig.~\ref{fig_PAs_vel_prof}, we compare the \lya\ emission profile extracted from the long-slit spectra obtained along PA = 72\degree\ (blue solid curve) and 350\degree\ (orange solid curve) after scaling the slit-loss corrected spectrum of PA = 350\degree\ to match with PA = 72\degree\  at the location of the red peak (shown by the thin dotted vertical line). While the profiles match nicely in the red part and in the core, we do see differences in the blue wing. The difference could be due to spatial differences in the H~{\sc i} absorption.

First we fit the emission profile along PA = 350\degree\ that was obtained during better seeing conditions. The obtained best fit values for log~[$N$(\hi) $\mathrm{cm^{-2}}$], \zabs\ and $b$  are 16.62\plm0.50, 3.13079\plm0.00005 and 62.2\plm14.5, respectively. The error on each parameter was calculated using 1000 randomly generated samples of the observed profile using errors on data points. The fit is shown by the red dash-dotted curve, and the unabsorbed profile is shown by solid red curve. Note that the large error in $N$(\hi) is mainly due to poor spectral resolution and a weak constraint on the emission line profile shape. The large b values compared to the thermal broadening of photoionized gas could reflect either large velocity dispersion or high saturation of the \lya\ absorption line that can appear unsaturated at the spectral resolution considered here. It is well known that the covering factor of the associated H~{\sc i} absorption need not be unity \citep[see, e.g.,][for example]{Fathivavsari2018}. In that case, the actual $N$(\hi) may be much higher than we recover.

We then fitted the spectrum obtained along PA = 72\degree, assuming the same emission profile as for PA = 350\degree\ and excluding the regions leftwards of the blue peak. The obtained fit is shown by the red dotted curve, and the best fit log~[$N$(\hi) $\mathrm{cm^{-2}}$], \zabs\ and $b$ are 18.99\plm0.05, 3.12998\plm0.00005 and 47.2\plm4.8 \kms, respectively. It is clear from this figure that the spectrum obtained along PA = 72\degree\ requires additional absorption components relative to the best fitted \lya\ emission profile obtained for PA = 350\degree. If we do not assume the same emission profile for the two PAs, then we can fit the \lya\ absorption with log~[$N$(\hi) $\mathrm{cm^{-2}}$], \zabs\ and $b$ equal to 15.15\plm0.79 cm$^{-2}$, 3.13032\plm0.0001 and 104.7\plm31.7, respectively. In the left panels of Fig.~\ref{fig_PAs_vel_prof}, we compare the profiles of He~{\sc ii} and C~{\sc iv}. The profiles obtained with PA = 350\degree\ have excess flux in the blue wing, as suggested by the \lya\ profile. Thus, allowing for variation in the emission profile along two PAs, results in similar absorption strength.

To check if absorption strength varies spatially, we also extract profiles along the dispersion axis around the center using spatial apertures of varying width, as shown in the lower left panel of Fig. \ref{fig_PAs_vel_prof}. Each profile has been scaled to match at the location of the blue peak (dotted vertical line). The figure shows that the absorption strength does not vary much with the spatial scale over which the spectra are extracted. We also see no significant change in the Ly$\alpha$ equivalent width with an increase in the width of the aperture over which spectrum is extracted. This means the absorbing region may be extended as much as the ``compact spatial'' component studied in section \ref{sub_extent}. However, any small spatial variations (over 2\arc\ scale) in the absorption profile might have gotten smoothed by the seeing smearing.

We further examine the spectral profiles extracted from the ``top" and ``bottom" regions shown in Fig. \ref{fig_vel_sections} to probe for the presence of extended absorption. Presence of weak absorption could be seen in these spectra as well, but constraining the column density is rather difficult due to poor SNR and spectral resolution. However, these observations are consistent with the spatially extended \lya\ absorbing region. Note in all these discussions, we assumed that the dip seen in \lya\ emission profile is due to absorption. The case would be strengthened if associated \civ\ absorption were detected using high SNR data, or if the \lya\ line profile were resolved at higher spectral resolution. Such evidence would rule out the possibility that some part or all of the \lya\ profile is being influenced by the radiative transfer effects that can generate complex profiles.

HzRGs are seen to show extended absorption up to \til 50 kpc, with absorption strength varying spatially as reported by VO97,  and with absorbers having column densities in the range 10$^{18}-10^{19.5}$\cms as measured from low resolution spectra  \citep[in some cases the actual column density is found to be lower when high resolution spectrum is used][]{jarvis2003}. For M1513-2524, we also see similar trend, however the column density for this source is lower than what has been found by VO97 using low-resolution data like ours. The absorber is blueshifted by \til 250-400 \kms with respect to the peak of the unabsorbed Gaussian profile of \lya\ emission along two PAs. In \cite{vanojik1997}, 60$\%$ of the sample showed \hi\ absorption and most of the absorbers were blue shifted within \til 250 \kms\ of \lya\ peak, which implied that these absorption were not likely to be caused by gaseous halos of neighbouring galaxies or by tidal remnants of interaction with nearby galaxies, but more likely associated with the intrinsic properties of HzRGs. We see a similar blueshift for M1513-2524 as well. However, VO97 found that most of the \hi\ absorption is found towards galaxies with smaller radio sizes, and only 25$\%$ of sources with $>50$ kpc radio sizes     showed absorption with only one source beyond 128 kpc radio size showing absorption (unlike in this case). Most of the properties of the absorber around M1513-2524 are consistent with VO97; however, M1513-2524 is the first of its kind in showing the presence of \hi\ absorption around a radio source extended by up to 184 kpc. We also obtained the uGMRT radio spectrum to detect presence of \hi\ 21-cm absorption. However, due to presence of RFI at the frequency of interest, we cannot say anything about possible \hi\ 21-cm absorption.

\section{Summary and Discussion}
\label{sec_summary}

In this paper, we have presented observations of M1513-2524, an optically faint source showing a very extended \lya\ nebula. For our study we have used long-slit spectra taken at two PAs along with narrow band imaging (both using RSS/SALT) and uGMRT radio maps at 1360 MHz and 420 MHz. Based on our analysis, together with existing archival data, we arrive at the following interesting results:

\begin{enumerate}
\item M1513-2524 is an extremely faint optical continuum source, not detected in any bands of the PS1 survey. The only emission lines that are detected are \lya, \nv, \civ\ and \heii. We used the non-resonant \heii\ line to estimate the redshift, \zem =3.1320\plm0.0003.
 
\item The total \lya\ luminosity of the nebula is (6.80\plm0.08)$\times 10^{44}$ \ergs, which is one among the highest luminosities detected so far. The \lya\ nebula extends out to \til\ 90 kpc for the 3\sig\ flux contour ($\mathrm{1.06 \times 10^{-18}}$ \ergscm) in the smoothed narrow band image. Based on the  similarity of line ratios with those of ``Non-\lya'' excess objects, we argue that the \lya\ emission (in particular from the compact region) in M1513-2524 is most likely powered by photoionization from the central AGN.
 
\item The 2D spectrum of the source clearly shows an associated \lya\ absorption demarcating the two spatial components, a ``compact'' and an ``extended'' component. The \civ\ and \heii\ profiles are significantly extended along PA =72\degree; however, the ``compact" component does not show clear extension along this PA. The PA=350\degree\ spectrum does not show significantly extended \civ\ and \heii\ emission; however it shows presence of asymmetry in the ``compact" component extending to \til\ 19 kpc. All these results emphasize the need for high SNR and high spatial resolution spectroscopy of M1513-2524.

\item We detect \lya\ absorption on top of the \lya\ emission blue shifted by 250-400 \kms\ with respect to the peak of \lya\ emission. Based on spectra taken along two position angles, we suggest that absorption may be extended. However, only high spatial resolution spectra would allow us to probe the spatial variations of the \lya\ absorption on top of the possible variations in the \lya\ emission line profile.

\item The high spatial resolution uGMRT radio maps confirm the presence of two radio lobes (separted by \til 184 kpc) associated with the source, and a weak core at the location of the WISE source at a 4$\sigma$ significance level. The \lya\ nebula is contained within the radio structure, with its outer regions more symmetric ($\alpha=0.84$) than its inner core region ($\alpha=0.79$) (see, e.g, Sec. \ref{sub_extent}). We find that the major axis of the isophotes for the core region is preferentially aligned close to the radio axis.
 
\item Kinematic analysis of the \lya\ emission shows that the central core region is more perturbed, with FWHM \til 1300 \kms, whereas the outer regions have FWHM \til 800-1000 \kms. Emission from the outer regions is redshifted with respect to the systemic redshift. It will be interesting to confirm the existence of these two distinct regions using IFS spectroscopy. Such high spatial resolution spectra together with high signal to noise radio images in the core regions will allow us to study any interaction between radio jet (or inner parts of the lobes) and the \lya\ emitting halo gas. 
 
\item We identify seven candidate \lya\ emitters around M1513-2524 having \lya\ luminosity greater than or equal to that of L* galaxies, while 0.2 galaxies are expected based on the observed \lya\ luminosity function. If these candidates are confirmed in followup spectroscopic observations, it will reveal that M1513-2524 may be part of a large proto-cluster of galaxies forming stars at a high rate. 

\end{enumerate}
In short, M1513-2524 has several properties similar to those of previously detected nebulae around HzRGs. However, there are several attributes that are rather unique to this source. First, both the radio power and \lya\ luminosity of M1513-2524 are among the highest known. Even more interesting is the detection of associated \lya\ absorption around a very large radio source, a configuration that is usually very rare. We have also identified seven potential \lya\ emitters around this radio source, which if confirmed could make M1513-2524 part of a proto-cluster of star forming galaxies. All these properties make M1513-2524 an ideal target for future integral field spectroscopic studies.
%
%
\section*{Acknowledgments}
We thank Dr. Montserrat Villar Martin for a very constructive report that has improved the presentation of this work.
Most of the observations reported in this paper were obtained with the Southern African Large Telescope (SALT). We thank the staff of the GMRT for wide band observations. GMRT is run by the National Centre for Radio Astrophysics of the Tata Institute of Fundamental Research. This work 
utilized the open source software packages \textsc{Astropy} \citep{astropy2}, \textsc{Numpy} \citep{numpy}, \textsc{Scipy} \citep{scipy}, \textsc{Matplotlib} \citep{matplotlib} and \textsc{Ipython} \citep{ipython}.
\section*{Data Availability}
Data used in this work are obtained using SALT. Raw data will become available for public use 1.5 years after the observing date at https://ssda.saao.ac.za/.
%
%
\def\aj{AJ}%
\def\actaa{Acta Astron.}%
\def\araa{ARA\&A}%
\def\apj{ApJ}%
\def\apjl{ApJ}%
\def\apjs{ApJS}%
\def\ao{Appl.~Opt.}%
\def\apss{Ap\&SS}%
\def\aap{A\&A}%
\def\aapr{A\&A~Rev.}%
\def\aaps{A\&AS}%
\def\azh{A$Z$h}%
\def\baas{BAAS}%
\def\bac{Bull. astr. Inst. Czechosl.}%
\def\caa{Chinese Astron. Astrophys.}%
\def\cjaa{Chinese J. Astron. Astrophys.}%
\def\icarus{Icarus}%
\def\jcap{J. Cosmology Astropart. Phys.}%
\def\jrasc{JRASC}%
\def\mnras{MNRAS}%
\def\memras{MmRAS}%
\def\na{New A}%
\def\nar{New A Rev.}%
\def\pasa{PASA}%
\def\pra{Phys.~Rev.~A}%
\def\prb{Phys.~Rev.~B}%
\def\prc{Phys.~Rev.~C}%
\def\prd{Phys.~Rev.~D}%
\def\pre{Phys.~Rev.~E}%
\def\prl{Phys.~Rev.~Lett.}%
\def\pasp{PASP}%
\def\pasj{PASJ}%
\def\qjras{QJRAS}%
\def\rmxaa{Rev. Mexicana Astron. Astrofis.}%
\def\skytel{S\&T}%
\def\solphys{Sol.~Phys.}%
\def\sovast{Soviet~Ast.}%
\def\ssr{Space~Sci.~Rev.}%
\def\zap{$Z$Ap}%
\def\nat{Nature}%
\def\iaucirc{IAU~Circ.}%
\def\aplett{Astrophys.~Lett.}%
\def\apspr{Astrophys.~Space~Phys.~Res.}%
\def\bain{Bull.~Astron.~Inst.~Netherlands}%
\def\fcp{Fund.~Cosmic~Phys.}%
\def\gca{Geochim.~Cosmochim.~Acta}%
\def\grl{Geophys.~Res.~Lett.}%
\def\jcp{J.~Chem.~Phys.}%
\def\jgr{J.~Geophys.~Res.}%
\def\jqsrt{J.~Quant.~Spec.~Radiat.~Transf.}%
\def\memsai{Mem.~Soc.~Astron.~Italiana}%
\def\nphysa{Nucl.~Phys.~A}%
\def\physrep{Phys.~Rep.}%
\def\physscr{Phys.~Scr}%
\def\planss{Planet.~Space~Sci.}%
\def\procspie{Proc.~SPIE}%
\let\astap=\aap
\let\apjlett=\apjl
\let\apjsupp=\apjs
\let\applopt=\ao

\defcitealias{gaikwad2016}{Paper-I}
\defcitealias{puchwein2015}{P15}	
\defcitealias{haardt2012}{HM12}	
\defcitealias{khaire2015a}{KS15}
\defcitealias{khaire2018a}{KS18}
\defcitealias{gaikwad2018}{G18}


 \bibliographystyle{mnras}
\bibliography{mybib}
\bsp
\label{lastpage}
\end{document}